\documentclass[superscriptaddress,twocolumn,showpacs,preprintnumbers,prb]{revtex4}
\usepackage{graphicx}
\usepackage{times}
\usepackage{epsfig}
\usepackage{color}
\usepackage{ulem}   
\begin{document}

\def\salto{\vskip 1cm} \def\lag{\langle} \def\rag{\rangle}

\newcommand{\redit}[1]{\textcolor{red}{#1}}
\newcommand{\blueit}[1]{\textcolor{blue}{#1}}
\newcommand{\magit}[1]{\textcolor{magenta}{#1}}

\newcommand{\MSTD} {Materials Science and Technology Division, Oak Ridge
National Laboratory, Oak Ridge, TN 37831, USA}
\newcommand{\CNMS} {Center for Nanophase Materials Sciences, Oak Ridge National
 Laboratory, Oak Ridge, TN 37831, USA}
\newcommand{\LLNL} {Lawrence Livermore National Laboratory, Livermore, CA
94550, USA}

\title{Self-healing diffusion quantum Monte Carlo algorithms: methods
  for direct reduction of the fermion sign error in electronic
  structure calculations}

\author{F. A. Reboredo}       \affiliation {\MSTD}
\author{R. Q. Hood}           \affiliation {\LLNL}
\author{P. R. C. Kent}        \affiliation {\CNMS}
\begin{abstract}
  We develop a formalism and present an algorithm for optimization of
  the trial wave-function used in fixed-node diffusion quantum Monte
  Carlo (DMC) methods. The formalism is based on the DMC mixed
  estimator of the ground-state probability density.  We take
  advantage of a basic property of the walker configuration
  distribution generated in a DMC calculation, to (i) project-out a
  multi-determinant expansion of the fixed-node ground-state wave
  function and (ii) to define a cost function that relates the
  fixed-node ground-state and the non-interacting trial wave
  functions.  We show that (a) locally smoothing out the kink of the
  fixed-node ground-state wave-function at the node generates a new
  trial wave-function with better nodal structure and (b) we argue
  that the noise in the fixed-node wave-function resulting from finite
  sampling plays a beneficial role, allowing the nodes to adjust
  towards the ones of the exact many-body ground state in a simulated
  annealing-like process.  Based on these principles, we propose a
  method to improve both single determinant and multi-determinant
  expansions of the trial wave-function. The method can be generalized
  to other wave-function forms such as pfaffians.  We test the method
  in a model system where benchmark configuration interaction
  calculations can be performed and most components of the Hamiltonian
  are evaluated analytically.  Comparing the DMC calculations with the
  exact solutions, we find that the trial wave-function is
  systematically improved.  The overlap of the optimized trial wave
  function and the exact ground state converges to 100\% even starting
  from wave-functions orthogonal to the exact ground state.
  Similarly, the DMC total energy and density converges to the exact
  solutions for the model.  In the optimization process we find an
  optimal non-interacting nodal potential of density-functional-like
  form whose existence was predicted in a previous publication [Phys.
  Rev.  B {\bf 77}, 245110 (2008)].  Tests of the method are extended
  to a model system with a conventional Coulomb interaction where we
  show we can obtain the exact Kohn-Sham effective potential from the
  DMC data.
\end{abstract}

\today

\maketitle
\section{Introduction}

In diffusion quantum Monte Carlo (DMC) a trial wave-function
is used to enforce both the antisymmetry of the electronic many-body
wave-function\cite{anderson79,ceperley80,reynolds82} and the nodal
structure of the solution. In highly correlated materials, the
accuracy of the trial wave-function becomes increasingly important and
determines the success or failure of the method. Indeed, concerns
about the fixed-node accuracy have tended to limit applications
of DMC to pre-transition metal elements. The discovery and
development of new methods to improve the trial
wave-functions, ideally without great computational expense, is
consequently highly desirable for almost all DMC
calculations.

In DMC calculations the trial wave-function $\Psi_T({\bf R})$ is
commonly a product of an antisymmetric function $\Phi_T({\bf R})$ and
a Jastrow factor $e^{J({\bf R})}$. Usually $\Phi_T({\bf R})$ is a
Slater determinant constructed with single particle Kohn-Sham orbitals
from density functional theory (DFT) or from other mean field
approaches such as Hartree-Fock. The Jastrow factor does not change
the nodes, but accelerates convergence and improves the algorithm's
numerical stability. The Jastrow factor is optimized in
a previous variational Monte Carlo (VMC) calculation. The DMC algorithm finds
the lowest energy of the set of all wave-functions that share the nodes of
$\Psi_T({\bf R})$. The exact ground-state energy will be obtained only if
the exact nodes are provided. Since any change to an antisymmetric
wave-function must result in a higher energy than the antisymmetric
ground state, the energy obtained with arbitrary nodes is an upper
bound to the exact ground-state energy.\cite{anderson79,reynolds82}
Only in small systems is it currently possible to improve the nodes
\cite{bajdich05,filippi00,umrigar07,rios06,luchow07} or even avoid the
trial wave-function approach altogether.\cite{ceperley84,kalos00,zhang91} For
small or weakly correlated systems, where other numerical approaches
can compete, the utility of DMC as a method depends crucially on
the accuracy of the trial wave-function. Multiple determinant,
pfaffian,\cite{bajdich05} and back-flow\cite{rios06} wave-functions
and geminal products\cite{beaudet} are increasingly popular due to
the improved accuracy.

To improve the DMC energy one must improve the nodal surface of the
trial wave-function. However, to our knowledge, all algorithms for
wave-function optimization are based on the VMC approach, with any
improvement in the DMC energy occurring only as a side-effect. The use
of VMC might be a limitation since VMC samples more frequently the
regions of the wave-function that have larger probability density and
are thus far from the nodes.\cite{luchow07} Accordingly, VMC based
optimization methods improve first the wave-function at regions which
are far from the nodes, while the nodes are only improved indirectly.
It has been found, however, that VMC based optimization methods, in
general, also improve the DMC energy.\cite{umrigar07,toulouse08}
Nevertheless, a direct optimization of the DMC energy is desirable,
and might have improved convergence properties compared to current
indirect approaches.

While it has been shown by us and others that, within the single
Slater determinant approach, the computational cost of an electronic
update step in the DMC algorithm can have an almost linear scaling
with the number of electrons,\cite{williamson,reboredo05,alfe04} the
use of these methods is limited if we do not find a better source
of trial wave-functions than those obtained from mean-field approaches
such as DFT. We recently showed\cite{rosetta} that Kohn-Sham DFT
wave-functions cannot
be expected to yield good nodes in general. As correlations
increase, Kohn-Sham DFT wave-functions can be bad sources of nodal
surfaces.\cite{rosetta} Indeed, we also found that as the size of the
system increases, the nodal error of DFT wave-functions might be of the
order of the triplet excitation energies, precluding the prediction of
accurate optical properties\cite{tiago08jpc} even for simple carbon
fullerenes. Accordingly, it is highly desirable to find a method to
(i) obtain trial wave-functions with accurate nodal structures, (ii)
retain the simplicity of a mean field approach, or (iii) use a
minimum number of Slater determinants i.e., the wave-functions are
compact and easily evaluated, (iv) directly optimize the nodes in DMC,
and (v) improve the nodal structure systematically independently of
the starting point. In this contribution we provide such a method.

In order to use DMC to find the best trial wave-function we overcome
two major obstacles: (i) obtain a representation of the fixed-node
ground-state DMC wave-function suitable for optimization of the
nodes, and (ii) find a method to keep the trial wave-function compact 
in large systems by minimizing the number of determinants.

This work is the natural continuation of a recent article
(Ref.~\onlinecite{rosetta}) where we proved the existence of an optimal
effective nodal potential for generating the orbitals in the determinants in
the trial wave-function used in DMC. While some details are rederived here, we
recommend reading Ref.~\onlinecite{rosetta} before this article. We previously
proved\cite{rosetta} that specific properties of the interacting
ground state can be retained via minimization of cost functions in the
set of pure-state non-interacting densities. Each cost function defines the
gradient of an effective non-interacting potential which is optimized
in a Newton-Raphson-like approach until the cost function reaches a
minimum. In this paper we take the next step: we use known properties
of the walker distribution function generated in a DMC run to
define a cost function relating the non-interacting wave-functions
with the fixed-node ground-state wave-function. This allows us to
obtain, for example, the Kohn-Sham potential or an effective nodal
potential from the DMC calculation. The method appears to be limited
by the quality of the fit, the statistics that one can collect in DMC and the
representability of the nodal surface, which becomes increasingly more
demanding as the number of electrons in the system increases.
Although this might limit the applicability of the
method to systems with small electron counts, we note that DMC is
readily parallelized with excellent scaling on modern computers. We
also expect that improved sampling and optimization schemes can be
constructed using the initial ideas and methods presented here and 
in Ref. \onlinecite{rosetta}. 

The remainder of this paper is organized as follows. In Section
\ref{sc:prove} we demonstrate that the nodes can be improved by locally
removing the kinks in the fixed-node ground state.  In Section
\ref{sc:multideterminants} we derive a formalism and a method to
obtain a multi-determinant expansion of the fixed-node ground-state
wave-function directly from a DMC run. For many applications, this
expansion may already be sufficient. In Section
\ref{sc:singledeterminant} we present a cost function that allows the
optimization of more compact trial wave-functions that match the fixed-node
ground state. A formalism for wave-function optimization based on an
effective DFT-like nodal potential is given. In Section \ref{sc:model}
we apply and compare these methods to a model system that can be
solved nearly analytically and demonstrate its convergence properties.
In Section \ref{sc:algorithm} we propose a general algorithm based on
the experience gathered solving the model. Finally in Section
\ref{sc:discussion} we summarize and discuss the prospects of this
method for application in large systems.

\section{Systematic reduction of the nodal error within DMC}
\label{sc:prove}
The importance sampling DMC algorithm, in the fixed-node
approximation, finds the lowest energy~\cite{fn:energy} $E^{DMC}_T$ among the set of
all wave-functions that
share the nodal surface $S_T({\bf R})$ where the trial wave-function 
$\Psi_T({\bf R}) = 0$ and changes sign. The symbol ${\bf R}$ denotes a point 
in the many-body $3 N$ dimensional space of electron coordinates.  We denote 
this wave-function $\Psi_{FN}({\bf R})$ as the fixed-node ground state. 
It can be shown that $\Psi_{FN}({\bf R})$ corresponds to the ground state 
of the interacting Hamiltonian containing an additional infinite external 
potential located at the nodes of $\Psi_{T}({\bf R})$.

\begin{figure}
\includegraphics[width=0.95\linewidth,clip=true]{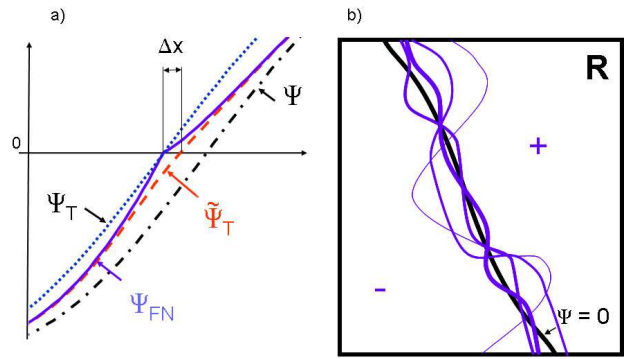}
\caption{a) (Color online) 
  Schematic representation of trial wave-function ($\Psi_{T}$, blue
  dots), fixed-node ground-state ($\Psi_{FN}$, purple continuous),
  ground-state ($\Psi$, black dash and dots), and new trial
  wave-function ($\tilde\Psi_T$, red dashed line) in the direction
  perpendicular to the nodal surface ($x$). We show that smoothing the
  kink in the fixed-node wave-function $\Psi_{FN}$ moves the nodes of
  $\tilde\Psi_T$ towards the nodes of the ground state $\Psi$ . b)
  Schematic representation of how the nodal surface evolves, shown
  with increasing purple line thickness, after each iteration in the
  algorithm. The noise introduced in the nodes by random fluctuations
  of the walkers is assumed to correct itself if the statistics is
  increased from one iteration to the next.}
\label{fg:nodes}
\end{figure}

The gradient of the fixed-node ground-state wave-function
$\Psi_{FN}({\bf R})$ can be discontinuous at the nodal surface
$S_T({\bf R})$.\cite{reynolds82} Indeed, if the nodes of the trial
wave-function do not correspond exactly to the nodes  $S({\bf R})$ of an 
eigenstate of the Hamiltonian, the Laplacian of the fixed-node
ground-state wave-function must have a $\delta({\bf R})$ contribution
at least on part of $S_T({\bf R})$. Otherwise, since the time independent Schr\"odinger
equation is satisfied elsewhere by $\Psi_{FN}({\bf R})$ with an energy $E^{DMC}_T$, without this
delta in the Laplacian at the nodal surface, $\Psi_{FN}({\bf R})$
would be an eigenstate of the Hamiltonian. This implies that the
gradient of $\Psi_{FN}({\bf R})$ must be discontinuous at least at one
point of $S_T({\bf R})$ if the nodal surface $S_T({\bf R}) \ne S({\bf
  R})$.

In Fig.~\ref{fg:nodes}a we show a schematic representation of the
trial wave-function $\Psi_T({\bf R})$, the ground-state wave-function
$\Psi({\bf R})$ and the fixed-node ground-state $\Psi_{FN}({\bf R})$.
In this section we show that when this kink in $\Psi_{FN}({\bf R})$ is
locally smoothed away as
\begin{eqnarray}
\label{eq:DMC}
\tilde\Psi_T({\bf R}) &=& \int {\bf dR^\prime} \Psi_{FN}({\bf R^\prime})
\tilde\delta \left({\bf R^\prime-R}  \right) \\
 &=&
 \int {\bf dR^\prime} \Psi_{FN}({\bf R+R^\prime})
\tilde\delta \left({\bf R^\prime}  \right), \nonumber
 \;
\end{eqnarray}
the nodes of the resulting functions improve for a broad class of
$\tilde\delta \left({\bf R-R^\prime}\right)$.

Provided that $\Psi_{FN}$ is an antisymmetric function with finite projection 
on the ground state $\Psi_0$, it has been shown\cite{anderson79,HLRbook} that $\Psi$, 
and its nodes converge to the exact ground state 
\begin{equation}
\label{eq:propagator}
\Psi = \lim_{t \rightarrow \infty} e^{-t(\hat{\mathcal{H}} - E_T)} \Psi_{FN}
\end{equation}
where $\hat\mathcal{H}$ is the Hamiltonian and $E_T$ is an estimate for
the ground-state energy. Setting $t=M\tau$ in Eq.~(\ref{eq:propagator})
yields the equivalent equation
\begin{equation}
\label{eq:propagator_m}
\Psi = \lim_{M \rightarrow \infty} \left( e^{-\tau(\mathcal{H} - E_T)} \right)^M \Psi_{FN} \; .
\end{equation}
In the limit of small $\tau$ a real-space linear-order expansion of 
$e^{-\tau(\mathcal{H} - E_T)}$ takes the form
\begin{eqnarray}
\label{eq:delta}
\tilde\delta \left({\bf R^\prime-R} \right) & = &   (2 \pi \tau)^{\frac{-3N}{2}} 
e^{-\tau \left(V({\bf R}) - E_T\right)}   
e^{\frac{-({\bf R - R^{\prime}})^2}{2 \tau}} \nonumber \\
& \simeq & \sum_n  e^{-\tau (E_n -E_T)} \left |\Psi_n \right >  \left < \Psi_n \right | 
\end{eqnarray}
where $V({\bf R})$ is the potential energy term (including interactions) in the Hamiltonian and
the $E_n $ are eigenvalues of the eigenvectors $ \Psi_n $.
Replacing the first line in Eq.~(\ref{eq:delta}) in
Eq.~(\ref{eq:DMC}) we obtain a function $\tilde\Psi_T({\bf R})$ that
has, by construction [see Eq. (\ref{eq:delta}) second line], an energy less than or equal to the energy of
$\Psi_{FN}({\bf R^\prime})$ (being equal for $S_T({\bf R}) = S({\bf
  R})$). This form of trial wave-function is similar to a shadow
wave-function.\cite{shadow,fn:purediff}
If we could evaluate Eq.~(\ref{eq:DMC})
analytically\cite{fn:bosons} and use the result $\tilde\Psi_T({\bf R})$ in a new DMC run,
we would obtain a new fixed-node ground-state wave-function with an even lower 
DMC energy. This implies that the nodes of $\tilde\Psi_T({\bf R})$ are 
better than the ones of $\Psi_T({\bf R})$.

Note that Eq.~(\ref{eq:delta}) tends to the Dirac $\delta$ function as
$\delta({\bf R})=(2 \pi \tau)^{-3N/2} e^{-({\bf R -
    R^{\prime}})^2/2\tau}$ for $\tau \rightarrow 0$. The factor
$e^{-\tau \left(V({\bf R}) - E_T\right)}$ in Eq.~(\ref{eq:delta}) does not alter the 
nodes: it is a positive scalar function (only acts as a branching term
in a one time step simulation). Accordingly, to linear order in $\tau$, only the Gaussian is required to 
improve the nodes.  In turn the Gaussian factor can be
replaced by any other approximation of the $\delta$ function as long as
it does the same to the nodes of $\Psi_{FN}({\bf R})$ as some Gaussian for small
$\tau$.  

In order to determine the class of smoothing functions that move the
node in Eq.~(\ref{eq:DMC}) as a Gaussian, we consider a patch $dS({\bf R})$
of the nodal surface $S_T({\bf R})$ centered at ${\bf R_S}$ with a diameter small
enough (so that it can be considered a flat hyper-plane) but much larger than
$\sqrt{\tau}$. The integration of the 3N dimensional Gaussian in the directions
of the hyper-plane leads to a one dimensional Gaussian
$G(x/\sqrt{\tau})= e^{ -x^2 /2\tau} / \sqrt{2 \pi \tau}$.
Any approximation of $\delta ({\bf R})$ after integration in $3N-1$
coordinates should result in a function $d(x)$ that can be rescaled and
translated to satisfy the following properties:
\begin{equation}
\label{eq:deltacond}
\int d(x) dx =1 \;\; \int x d(x) dx =0 \; \text{and} \; \int x^2 d(x) dx =1 \; .
\end{equation}

In the immediate vicinity of ${\bf R_S}$, the function $\Psi_{FN}({\bf R})$
depends only on the coordinate in the direction
normal to the surface ${\bf n_S}$ defined as $x={\bf (R -R_S) \cdot n_S}=
{\bf \Delta R \cdot n_S}$. For $x \rightarrow 0 $ we can approximate
\begin{eqnarray}
\label{eq:appsi}
\Psi_{FN}({\bf R_S+ \Delta R} ) \simeq \Psi_{FN}({\bf R_S}) + c_1 x + k_1 |x| && + \nonumber \\ 
+ c_2 x^2 + k_2(x-|x|)^2 + O(x^3)
\end{eqnarray}
and
\begin{eqnarray}
\label{eq:apderpsi}
\frac{d}{dx}\Psi_{FN}({\bf R_S+\Delta R} ) \simeq c_1 + k_1 \mathrm{sign}(x) + &&
\\ +
2 c_2 x +  4\; k_2 (x - |x|)+O(x^2). \nonumber 
\end{eqnarray}
In Eq.~(\ref{eq:appsi}) the wave-function is expanded as a combination of a
smooth function (with coefficients $c_1$ and $c_2$) plus a kink
($k_1$ and $k_2$). Replacing Eq.~(\ref{eq:appsi}) and Eq.~(\ref{eq:apderpsi})
into Eq.~(\ref{eq:DMC}) and replacing the Gaussian by a generic approximation
of $\delta(x)=d(x/\sqrt{\tau})/\sqrt{\tau}$ we get:
\begin{equation}
\label{eq:val}
\tilde\Psi_T({\bf R_S}) = k_1 A[d] \sqrt{\tau}+O(\tau)
\end{equation}
and the first derivative
\begin{equation}
\label{eq:der}
\frac{d}{dx} \tilde\Psi_T({\bf R_S}) = c_1+ k_1 S[d]+ 4 k_2 A[d] \sqrt{\tau} +
O(\tau)
\end{equation}
where $A[d]=\int |x| d(x) dx$ and $S[d]=\int \mathrm{sign}(x) d(x) dx$. Note
that if $d(x)$ has the Gaussian form $A[G]=\sqrt{2/\pi}>0$ and $S[G]=0$. Using
Eqs.~(\ref{eq:val}) and (\ref{eq:der}) we
can estimate the displacement of the node to be
\begin{equation}
\label{eq:newnode}
\Delta x \simeq  -\frac{k_1 A[d]}{c_1 + k_1 S[d]} \sqrt{\tau} + O(\tau).
\end{equation}
Therefore, for any symmetric approximation of the $\delta$ function
$S[d]=0$, provided that  $A[d]>0$, one can obtain the same displacement in the
node as a Gaussian with $\tau^\prime = \pi A[d]^2 \tau/2$. For a non-symmetric
$d(x)$, the node will move in the same direction as long as the sign in the
denominator of Eq.~(\ref{eq:newnode}) does not change. However, a uniform
rescaling of $\tau$ to match the Gaussian form will no longer be possible.
That means that the node will move faster towards the exact node in some
regions of the surface than in others.

Thus, as long as the approximation of the delta used for smoothing is
a function of the distance only, with $A[d]>0$, one can find some
Gaussian that moves the node in the same way for every patch $dS({\bf
  R})$.  This movement corresponds to a better node.  The restrictions
in $d(x)$ can be alleviated by using a repeated convolution.  Using
the central limit theorem it can be shown that a recursive convolution
of any approximation of $d(x)$ tends to a Gaussian as long as the
Taylor expansion of its Fourier transform exists.  Thus if the shape
of $d(x)$ is not known the method would be more stable if it is
applied sequentially.

In section \ref{sc:multideterminants} we will use a smoothing function
of the form
\begin{equation}
\label{eq:deltaexp}
\tilde \delta
\left({\bf R,R^\prime} \right) =
\sum_n^{\sim}
\Phi_n({\bf R}) \Phi_n^*({\bf R^{\prime}}),
\end{equation}
where the $\Phi_n({\bf R})$ are continuous functions without kinks
forming a complete basis and the ``$\sim$'' in $\sum_n^{\sim}$ means
that only some elements are included in the sum (with a criterion
described below). If the $\Phi_n({\bf R})$ in Eq.~(\ref{eq:deltaexp})
are obtained from a non-interacting problem and the criterion for
truncation is an energy cutoff, it can be shown that the resulting
function is only a function of the distance ${\bf (R-R^\prime)}^2$.
Since in that limit only plane waves of large energy are added to
Eq.~(\ref{eq:deltaexp}) and all the lower plane waves are included in
the lower energy components, the basis can be transformed with a
unitary transformation into a plane-wave basis with a spherical cutoff
in reciprocal space.  If there is the same number of plane waves in
any direction the results of Eq.~(\ref{eq:deltaexp}) only depend on
the distance which implies that $S[d]=0$.

Since we restrict the sum in Eq.~(\ref{eq:deltaexp}) to fermionic
antisymmetric $\Phi_n({\bf R})$, Eq.~(\ref{eq:deltaexp}) expands an
antisymmetrized delta\cite{fn:bosons}.  This form projects out any non-fermionic
component introduced in the wave-function along the DMC algorithm as
in the A-function approach used by Bianchi and collaborators.\cite{bianchi93}

In Section \ref{sc:singledeterminant} we propose a simple
interpolation scheme to smooth the node where the expansion used in
Eq.~(\ref{eq:deltaexp}) is not taken to the high energy cutoff limit.
The fact that these smoothing methods work in practice suggests that
the conditions to improve the nodes are extended beyond the exact
equivalence to a Gaussian form.

Note that a discontinuity of the gradient of the fixed-node
wave-function $\Psi_{FN}({\bf R})$ at the node implies\cite{fn:purediff} that, if
walkers are distributed according to $\Psi_{FN}({\bf R})$ with the
sign (or phase) of $\Psi_{FN}({\bf R})$, there will be more walkers in
the vicinity of one side of the nodal surface than on the other.
Accordingly, if these walkers are released in a pure diffusion
algorithm,\cite{HLRbook} for $\tau \rightarrow 0$ they will cross, on
average, more from one side of the nodal surface than from the other.
The nodes defined by the population of these signed
walkers\cite{HLRbook} would move in the same direction that would
result from smoothing the kink in $\Psi_{FN}({\bf R})$ provided the
time step is short enough and kinetic energy term in the Green's
function [Eq.~(\ref{eq:delta})] is dominant.  Consequently, the nodes
can be improved by moving them in the direction of least ``walker
pressure'' within a pure diffusion approach.

Any method to obtain $\Psi_{FN}({\bf R})$ from the walker
distribution in a DMC run\cite{bianchi96} will carry the error of
statistical fluctuations from using a finite sample of walkers. Even if the
$\Psi_{FN}({\bf R})$ is forced to remain antisymmetric,\cite{bianchi96}  the
nodes might move in the wrong direction because of these fluctuations.
We assume the method is robust against these random fluctuations when
applied recursively, and can form the basis of an optimization process
to improve the trial wave-function. Note that if incorrect
fluctuations increase the kink in $\Psi_{FN}({\bf R})$ at the node,
the probability to sample the correct fixed-node wave-function will
remain higher and also the probability to move the node in the correct
direction in successive iterations.  Conversely, fluctuations that
correctly improve the nodes will be reinforced\cite{fn:reinforced} in successive
iterations. Since these fluctuations are reduced when the statistical sampling
is improved, the nodal surfaces will converge to the true nodes if the
statistics is improved from one iteration to the next
(Fig.~\ref{fg:nodes}b).  Note that we do not claim that this process
is necessarily the most efficient optimization approach: more
sophisticated iterative methods and optimization algorithms are
clearly possible.

Summarizing, we should be able to improve the nodes systematically
provided we can obtain {\it the anti-symmetric function}
$\Psi_{FN}({\bf R})$ from the walker configurations (probability
distribution) of a DMC calculation after convolution with a smoothing
function.\cite{fn:relnode}

\section{Determination of the fixed-node ground-state wave-function
from the DMC probability distribution \label{sc:multideterminants}}

\subsection{Sampling the fixed-node ground-state wave-function}

The distribution function of the walkers in an importance sampling DMC
algorithm is given by:\cite{ceperley80}
\begin{eqnarray}
\label{eq:eta}
f({\bf R}) & = & \Psi_T^*({\bf R})\Psi_{FN}({\bf R}) \\
           & = & \lim_{N_c \rightarrow \infty} \frac{1}{N_c} \sum_i^{N_c} \delta \left({\bf R-R}_i \right) \nonumber
\end{eqnarray}
where $\Psi_T({\bf R})$ typically has the Slater-Jastrow form
\begin{equation}
\label{eq:psit}
\Psi_T({\bf R}) = e^{J({\bf R})} \Phi_T({\bf R});
\end{equation}
in which $\Phi_T({\bf R})$ consists of a single determinant for each
electronic spin component composed of single-particle orbitals. The
results of this paper are also valid if $\Phi_T({\bf R})$ has a more
general form such as consisting of multi-determinant expansions for
each spin component and/or containing back-flow or two-particle
pfaffians.
The ${\bf R}_i$ in Eq. (\ref{eq:eta}) correspond to the positions of an
equilibrated ensemble of $N_c$ configurations in a DMC algorithm  (we have set
the weights equal to one for simplicity).  

We note that $\Psi_{FN}({\bf R})$ in Eq.~(\ref{eq:eta}) can be
rewritten as an antisymmetric function times the Jastrow factor
$e^{J({\bf R})}$ as
\begin{eqnarray}
\label{eq:expan}
\Psi_{FN}({\bf R}) & = &  e^{J({\bf R})}  e^{-J({\bf R})} \Psi_{FN}({\bf R})
\nonumber \\
&=&  e^{J({\bf R})} \sum_n \lambda_n \langle{\bf R}|
(\prod c^{\dag} \prod c) |\Phi_T \rangle \nonumber \\
 &=& e^{J({\bf R})} \sum_n \lambda_n \Phi_n ({\bf R})
\end{eqnarray}
where $\sum \lambda_n (\prod c^{\dag} \prod c) |\Phi_T \rangle$ is a
complete configuration interaction (CI) expansion in the basis of
electron-hole pairs .  Accordingly, in Eq.~(\ref{eq:expan}) the
$\Phi_n ({\bf R})$ are Slater determinants or pfaffians\cite{bajdich05}
obtained from replacing in $\Phi_T ({\bf R})$ some of the occupied
$\phi_{\nu}$ single particle functions by unoccupied $\phi_{n}$ functions,
accordingly
$\int {\bf d R} \Phi_n^* ({\bf R}) \Phi_m ({\bf R}) = \delta_{n,m}$.

In practice, the CI expansion can be truncated retaining, for example,
only the $\Phi_m ({\bf R}) $ with a non-interacting
energy below a given energy cutoff. The CI expansion in principle
consists of all single, double, triple, quadruple and higher-order
excitations. By analogy with conventional CI calculations, the
higher-order excitations are expected to contribute less to the
wave-function than low-order excitations. As the kinetic energy of
higher-order excitations increases as compared with the interaction,
their contribution to the ground-state wave-function decreases.

While a Jastrow factor $e^{J({\bf R})}$ is not formally required in a
complete expansion of the wave-function in Eq.~(\ref{eq:expan}), it is
believed that the introduction of a Jastrow factor limits the number
of coefficients required in the multi-determinant expansion, due in
part to the more efficient description of the electron-electron cusp.
For some applications it may be desirable to {\it not} employ a
Jastrow factor, since the extracted wave-function may be more easily
used in later analysis.

Replacing Eq.~(\ref{eq:expan}) and Eq.~(\ref{eq:psit}) in Eq.~(\ref{eq:eta})
we obtain
\begin{equation}
\label{eq:expaneta}
f({\bf R})= e^{2J({\bf R})}\Phi_T^* ({\bf R}) \sum_n \lambda_n \Phi_n ({\bf R}) .
\end{equation}
Borrowing a method from Optimized Effective Potentials (OEP) we
define the following projectors:\cite{us,note}
\begin{equation}
\label{eq:xi}
\xi_n({\bf R})= e^{-2J({\bf R})} \frac {\Phi_n ({\bf R})} { \Phi_T ({\bf R})} .
\end{equation}
Note that the projectors $\xi_n({\bf R})$ are symmetric (bosonic) functions.\cite{fn:bosons}
Replacing $f({\bf R})$ by Eq.~(\ref{eq:expaneta}), using the definition
of $\xi_n({\bf R})$ [Eq.~(\ref{eq:xi})] and the orthogonality
condition it can be demonstrated that
\begin{equation}
\label{eq:coeff}
\int {\bf dR}  f({\bf R}) \xi_n^*({\bf R}) = \lambda_n \; .
\end{equation}
Thus, the coefficients of the multi-determinant expansion
Eq.~(\ref{eq:expan}) of the fixed-node DMC ground-state wave-function
can be estimated directly as a sum over the total number of walkers
$N_c$ along the DMC random walk, using the second line of Eq.
(\ref{eq:eta}), as
\begin{equation}
\label{eq:sampcoeff}
\langle \lambda_n \rangle = \frac{1}{N_c} \sum_{i=1}^{N_c}
\xi_n^*({\bf R}_i) \;
\gamma({\bf R}_i)
\end{equation}
where
\begin{equation}
\label{eq:gamma}
 \gamma ({\bf R}_i)= \frac{-1 + \sqrt{1 + 2 |{\bf v}|^2 \tau}}
{|{\bf v} |^2  \tau} \text{ with }
{\bf v} = \frac{\nabla \Psi_T({\bf R}_i)} {\Psi_T({\bf R}_i)}.
\end{equation}
For convenience we divided by the number of walkers $N_c$ in
Eqs.~(\ref{eq:eta}) and~({\ref{eq:sampcoeff}) since the normalization constant
of $\Psi_{FN}({\bf R})$ and the corresponding coefficients
$\lambda_n$ is arbitrary. The factor $\gamma ({\bf R}_i)$ in
Eq.~(\ref{eq:sampcoeff}) is a time step, $\tau$, correction
derived following Ref.~\onlinecite{umrigar93} that corrects
the divergences of the projectors $\xi_n({\bf R}_i)$ at the
nodes. This correction is not always applied to estimators (e.g.
the local energy) but we find that it reduces the
error of the wave-function coefficients. For an uncorrelated
sample of walker configurations the error bar of the multi-determinant
expansion can be determined from
\begin{eqnarray}
\label{eq:errorbar}
\langle \lambda_n^2 \rangle &=& \frac{1}{N_c} \sum_{i=1}^{N_c}
|\xi_n({\bf R}_i)|^2
\gamma({\bf R}_i)^2 \\
\langle{\sigma_n}\rangle & = & \sqrt{\frac{\langle\lambda_n\rangle^2-
\langle\lambda_n^2\rangle}
{N_c}} \nonumber \\
\lambda_n & \simeq & \langle\lambda_n\rangle
\pm \frac{\langle{\sigma_n}\rangle}{\sqrt{N_c-1}}. \nonumber
\end{eqnarray}
As $N_c \rightarrow \infty$ in Eq.~(\ref{eq:errorbar}) the error bar
in the multi-determinant coefficients $\lambda_n$ goes to zero.  As
usual, the error bars can be used to monitor convergence of the
calculation. While the eventual goal is to obtain small error bars, we
found in practice it is better to start with $N_c$ small and then to
slowly increase it with each iteration as the trial wave-function
improves (see below).

By substituting Eqs.~(\ref{eq:eta}), (\ref{eq:psit}), and (\ref{eq:xi})
into Eq.~(\ref{eq:coeff}) and defining the fixed-node function
$\Phi_{FN}$ in terms of the trial function Jastrow and the fixed-node
wave-function $\Psi_{FN}$
\begin{equation}
\Psi_{FN}({\bf R}) = e^{J({\bf R})} \Phi_{FN}({\bf R})
\end{equation}
one can obtain this expression
\begin{equation}
\label{eq:lambda_expression}
\lambda_n = \int {\bf dR} \Phi_n^*({\bf R}) \Phi_{FN}({\bf R})
\end{equation}
for $\lambda_n$. We define $\tilde\Psi_T({\bf R})$ to be the truncated
expansion (denoted using $\sim$) of Eq.~(\ref{eq:expan})
\begin{equation}
\label{eq:psit_trunc}
\tilde\Psi_T({\bf R})=e^{J({\bf R})} \sum_n^{\sim} \lambda_n \Phi_n ({\bf R}) \; .
\end{equation}
Substituting Eq.~(\ref{eq:lambda_expression}) into Eq.~(\ref{eq:psit_trunc})
yields the equation
\begin{equation}
\label{eq:psit_trunc_delta}
\tilde\Psi_T({\bf R})=e^{J({\bf R})} \int {\bf dR^\prime} \left[ \sum_n^{\sim}
\Phi_n({\bf R})
\Phi_n^*({\bf R}^\prime) \right] \Phi_{FN} ({\bf R}^\prime) \; .
\end{equation}
In Section \ref{sc:prove} we showed that the appearance of a smoothing
function of the form of Eq.~(\ref{eq:deltaexp}) as in the term in
brackets in Eq.~(\ref{eq:psit_trunc_delta}) will smooth the nodes of
$\Phi_{FN} ({\bf R}^\prime)$ yielding {\it better nodes} for
$\tilde\Psi_T({\bf R})$. Since the $\Phi_n ({\bf R})$ are selected to
be eigenvectors of a non-interacting problem, highly localized
features of $\Phi_{FN}({\bf R})$ would require components with high
eigenvalues.  At the same time, resolving those details would require
a large number of configurations to improve the statistics.
Accordingly, we truncate the expansion in Eq.~(\ref{eq:psit_trunc}) to
the coefficients with relative errors smaller than 25\%. Note that as
the statistics is improved, the error bars diminishes, the number of
functions retained in Eq.~(\ref{eq:deltaexp}) increases and so does
the localization of $\tilde\delta\left({\bf R,R^\prime} \right)$. Thus
the conditions to improve the nodes systematically as described in
Section \ref{sc:prove} are reached as the statistics improves.

\subsection{Sampling the Jastrow factor}
Instead of expressing $\Psi_{FN}({\bf R})$ as a product of
the same Jastrow factor used in $\Psi_T({\bf R})$ times a different
multi-determinant expansion, one can choose to optimize the Jastrow factor
while using the same antisymmetric function $\Phi_T({\bf R})$. It is
easy to show that there is a symmetric bosonic factor that turns
$\Phi_T({\bf R})$ into $\Psi_{FN}({\bf R})$ which is formally given by
\begin{equation}
\label{eq:getJastrow}
e^{\tilde{J}({\bf R})} = \frac{ \Psi_{FN}({\bf R})}{ \Phi_T({\bf R})} \; .
\end{equation}
Replacing Eq.~(\ref{eq:expan}) in Eq.~(\ref{eq:getJastrow}) we find
\begin{eqnarray}
\label{eq:giveJastrow}
e^{\tilde{J}({\bf R})} & = &  e^{J({\bf R})} \sum_n \lambda_n
\frac {\Phi_n ({\bf R})} { \Phi_T ({\bf R})} \nonumber \\
 &=&  e^{3 J({\bf R})} \sum_n \lambda_n \xi_n({\bf R})  \;. 
\end{eqnarray}
Note that the product $e^{\tilde{J}({\bf R})} \Phi_T ({\bf R})$ yields
Eq.~(\ref{eq:expan}). While this shows that the projectors $\xi_n({\bf
  R})$ could be used to improve the Jastrow factor, since they diverge
for $\Phi_T({\bf R}) \rightarrow 0$, it is necessary to fit instead a
continuous functional form using values away from the nodes where
truncation and sampling errors play a dominant role (see Section
\ref{sc:singledeterminant}).

Updating the multi-determinant expansion of the antisymmetric part of
the new trial wave-function, see Eq.~(\ref{eq:psit_trunc}), alters the
nodes because (i) the expansion is truncated and (ii) the coefficients
of the multi-determinant expansion have a random error due to finite
sampling in Eq.~(\ref{eq:sampcoeff}). On the other hand, updating the
Jastrow factor, see Eq.~(\ref{eq:giveJastrow}), keeps the nodes fixed
but reduces the number of determinants required and the overall
computational cost.  There is a compromise between accuracy and
speed.\cite{reynolds82} A very good wave-function might have a very
small variance in the local energy, but if it is expensive to evaluate
one might obtain the same statistical error in less wall-clock time
with a faster lower quality wave-function. In an ideal case, if the
nodes are $v$-representable (see below and Ref.~\onlinecite{rosetta})
only a single determinant is required to describe the fixed-node
ground-state wave-function to sufficient accuracy. In practice, the
form of the Jastrow factor $e^{\tilde{J}({\bf R})}$ is unknown, while
an infinite multi-determinant expansion is infeasible. This implies
that both the factors in Eq.~(\ref{eq:expan}) are required in general;
an efficient scheme will optimize both the Jastrow factor and
determinantal part of the wave-function. Particularly for the case of
a metallic system, the cost of a multi-determinant expansion might be
prohibitive due to the large number of low-energy excitations. In this
case it might be preferable to concentrate on an optimized Jastrow
factor.\cite{wood06}

\subsection{A simple self-healing DMC algorithm}
We have formulated, for small systems, a working iterative algorithm
based on a multi-determinant or multi-pfaffian expansion of the
fixed-node ground-state wave-function. In this algorithm the
calculated coefficients Eq.~(\ref{eq:sampcoeff}) of the expansion are
used to form a new trial wave-function defined by
Eq.~(\ref{eq:psit_trunc}).  Initially the statistical errors present
in $\lambda_n$ due to finite sampling appear to have a beneficial
role, particularly when the initial trial wave-function has poor
nodes. Note that in the limit of an infinite number of determinants in
Eq.~(\ref{eq:psit_trunc}) with no statistical sampling errors in
$\lambda_n$ the trial wave-function would exactly reproduce the
fixed-node wave-function, and an iterative improvement of the nodes
would not be possible. Statistical fluctuations in the coefficients
$\lambda_n$ allow the nodes to move. In the next iteration regions
near beneficial fluctuations are revisited by walkers while
bad statistically insignificant fluctuations tend not to propagate or
grow. This stability against random noise appears to be valid in
practice.  Thus, the statistical error in the coefficients plays the
role of a random thermal fluctuation in a simulated annealing
algorithm.\cite{correa05} It is ironic and remarkable that random
errors can be used to eliminate systematic errors.

While it is relatively economical to calculate a large number of
multi-determinants every autocorrelation length, as more determinants
are included in the trial wave-function each time step of the DMC
calculation becomes more demanding. Accordingly, for large or
continuum systems a method to minimize the number of determinants used
to represent a given nodal surface is required. This is described in
the next section.

\section{Derivation of the best nodal-effective potential from DMC}
  \label{sc:singledeterminant}}
While a working multi-determinant algorithm can be constructed on the
basis of the multi-determinant expansion of the previous section, a
significant step forward can be taken using the theory developed in
Ref.~\onlinecite{rosetta} and taking advantage of Eq.~(\ref{eq:eta})
to construct a new trial wave-function that can be evaluated more
efficiently than the multi-determinant expansion Eq.
(\ref{eq:psit_trunc}). This method will be most effective when the
initial single particle orbitals involved in $\Phi_T({\bf R})$ are
poor, e.g. if the system is strongly correlated.

\subsection{A cost function for the  DMC algorithm}
\label{ssc:costfunction}
Given a probability density $p({\bf R})$ and a binned statistical sample of
$N_c$ configurations of the random variable ${\bf R}$, we can
define a new random variable
\begin{equation}
\label{eq:chi2}
\chi^2= \sum_i^M
\frac{\left[ n_i - N_c \Omega_i \bar{p}({\bf R}_i) \right ]^2}
{N_c \Omega_i \bar{p}({\bf R}_i)}
\end{equation}
which is distributed by the Chi-squared distribution function.\cite{HLRbook}
In Eq.~(\ref{eq:chi2}) $\Omega_i$ is the volume of the bin $i$, with $n_i$
configuration counts, $\bar{p}({\bf R}_i)$ is the average of $p({\bf R})$
in $\Omega_i$ and $M$ is the number of bins. 

Each term in Eq.~(\ref{eq:chi2}) is the square deviation of $n_i$
divided by the expectation value of the mean. In the limit of large
counts the square of the mean is expected to be equal to the square
deviation for the Poisson distribution of counts in a bin.
Accordingly, in $\chi^2$ relative deviations from the mean have the
same impact independently of the absolute value of the probability
density. We will take advantage of this property to replace a
wave-function difficult to evaluate Eq.~(\ref{eq:eta}) by a simpler
approximate one that retains key properties.  Setting $n_i= N_c
\Omega_i \bar{q}({\bf R}_i)$ in Eq.~(\ref{eq:chi2}), dividing by $N_c$
taking the limit $M \rightarrow \infty$, and using the mean value
theorem, we find a cost function to compare two continuous
distribution functions:
\begin{equation}
\label{eq:p.vs.q}
K_{pq}= \int {\bf dR}
\frac{\left[q({\bf R}) - p({\bf R}) \right ]^2}
{p({\bf R})} \; .
\end{equation}

We showed in Ref.~\onlinecite{rosetta} that if we wish to preserve
properties, other than the density, cost functions can be
defined relating the many-body ground-state $\Psi({\bf R})$ with a
non-interacting wave-function $\Phi_{T}({\bf R})$.
The walker distribution function\cite{ceperley80} given by Eq.
(\ref{eq:eta}) allows us to construct several cost functions relating
the wave-function to optimize with the exact fixed-node ground-state
$\Psi_{FN}({\bf R})$. Using Eq.~(\ref{eq:p.vs.q}) as a guide, we
propose the following expression:
\begin{eqnarray}
\label{eq:KDMC}
K_{DMC} & = & \int {\bf dR}
\frac{\left | \mu \; \Psi_T({\bf R})\tilde{\Psi}_T({\bf R}) - f({\bf R}) \right |^2}
{\left | \mu \; \Psi_T({\bf R}) \tilde{\Psi}_T({\bf R}) \right |} \times
\nonumber \\
&  & \theta \left( f({\bf R})  - \eta \right ) ,
\end{eqnarray}
where $\tilde{\Psi}_T({\bf R})$ is a trial wave-function to be
optimized, $\mu = \left[\int \Psi_T({\bf R})\tilde{\Psi}_T({\bf R})
  {\bf dR} \right]^{-1} $, $f({\bf R})$ is given by
Eq.~(\ref{eq:expaneta}) with coefficients obtained from a previous DMC
run using Eq.  (\ref{eq:sampcoeff}), $\theta(x)$ is the Heaviside
function, and $\eta$ is a small positive n umber.  Note in
Eq.~(\ref{eq:KDMC}) that the first factor vanishes when $
\tilde{\Psi}_T({\bf R})\rightarrow \Psi_{FN}({\bf R})$.  Indeed, if
$\tilde{\Psi}_T({\bf R})$ is constrained to have the nodal surface
$S_T({\bf R})$ and the sign (or phase) of $\Psi_T({\bf R})$, the
integral of the first factor in Eq.~(\ref{eq:KDMC}) measures the
probability that the distribution of a given ensemble of walkers
$f({\bf R})$ corresponds to the distribution\cite{HLRbook}
\begin{equation}
\alpha({\bf R}) = \mu \tilde{\Psi}_T({\bf R})\Psi_{T}({\bf R}) \; .
\end{equation}

In Eq.~(\ref{eq:KDMC}), we add an absolute value function in the
denominator of the first factor and a Heaviside function in order
to extend the set of $\tilde{\Psi}_T({\bf R})$ where the cost function
can be evaluated beyond the fixed-node space. Note that, since
$f({\bf R}) > 0$, while negative values for $\alpha({\bf R})$ are
allowed, they are penalized in the numerator more than positive values.
In Eq. ~(\ref{eq:KDMC}), we add $\mu$ to enforce 
$\int \alpha({\bf R})  {\bf dR } = \int f({\bf R}) {\bf dR } $ for 
any $\tilde{\Psi}_T({\bf R}) $.
In Eq.~(\ref{eq:KDMC}) the nodes of $\tilde{\Psi}_T({\bf R})$ can
move within a distance [which depends on $\eta$ and $f({\bf R}$)]
around $S_T({\bf R})$. Otherwise, if the zeros of the numerator and
denominator of Eq.~(\ref{eq:KDMC}) do not match, the value of the cost
function would rise to infinity. An additional effect of $\theta$ is that
any kink of $\Psi_{FN}({\bf R})$ at the node is not enforced by the
cost function in $\tilde{\Psi}_T({\bf R})$.  Since
$\tilde{\Psi}_T({\bf R})$ will be obtained from the minimum energy
solution of a non-interacting problem\cite{rosetta} and departures at
the node are not penalized, it will interpolate smoothly avoiding a
kink. Note that we can choose alternative cost function forms.  For
example, we can replace the denominator in Eq.~(\ref{eq:KDMC}) by
$f({\bf R}$). This choice would simplify the derivatives of the cost
function but it has a couple of disadvantages: First $f({\bf R}) $ is
expected to be a very noisy function when its magnitude is small,
while the product of non-interacting $v$-representable wave-functions
$\alpha({\bf R}) = \mu \; \tilde{\Psi}_T({\bf R})\Psi_{T}({\bf R})$ is
expected to be smooth (see \ref{ssc:represent}) .  We choose not to
amplify the noise of $f({\bf R}) $ in the denominator.  Second, in
Eq.~(\ref{eq:KDMC}) a small number for $\alpha({\bf R})$ outside the
window defined by the Heaviside function is highly penalized which
confines the node of $ \tilde{\Psi}_T({\bf R})$ to remain inside the
window where the Heaviside function is zero.

\subsection{Representability of the nodal surface}
\label{ssc:represent}
Given an interaction in a many-body system, the Hohenberg-Kohn
theorem\cite{hohenberg} establishes a functional correspondence
between electronic densities $\rho({\bf r})$, external potentials $V({\bf r})$,
and ground-state wave-functions $\Psi({\bf R})$. The subset of densities
$\rho({\bf r})$ corresponding to a ground state of an interacting system under
an external potential $V({\bf r})$ are denoted as {\it pure state}
$v$-representable.\cite{parr} A non-interacting {\it pure state}
$v$-representable density is given instead by $\bar\rho({\bf r})=
\sum_{\nu} |\phi_{\nu} \left({\bf r} \right)|^2$ where $\phi_{\nu
} \left({\bf r} \right)$ are Kohn-Sham-like\cite{kohn} single particle
orbitals, or eigenvectors, of the single-particle Hamiltonian:
\begin{equation}
\label{eq:Heff}
\left[ -\frac{1}{2}{\bf \nabla}^2
+\bar V \left({\bf r}  \right) \right] \phi_{\nu}\left({\bf r} \right)
=\varepsilon_{\nu} \phi_{\nu} \left( {\bf r}\right),
\end{equation}
where $\bar V \left({\bf r} \right)$ is an effective single particle
potential. The lowest energy Slater determinant constructed with the
solution of Eq.~(\ref{eq:Heff}) is a many-body non-interacting ground
state.  For simplicity we denote those quantities that are
simultaneously interacting and non-interacting $v$-representable as
simply {\it $v$-representable}. In addition, certain quantities can be
{\it multi-determinant} $v$-representable, meaning that they can be
represented by a finite multi-determinant expansion constructed with
the solutions of Eq.~(\ref{eq:Heff}). Since, the ground-state density
$\rho({\bf r})$ determines the ground-state wave-function $\Psi({\bf
  R})$,\cite{hohenberg} $\rho({\bf r})$ defines also the points ${\bf
  R}$ of the nodal surface $S({\bf R})$ where $\Psi({\bf R}) = 0$. The
nodes of the trial wave-function, instead, are by construction those
of $\Phi_T({\bf R})$ (non-interacting $v$-representable in the single
determinant case). The exact nodes $S({\bf R})$ may or may not be
representable in this manner.\cite{rosetta}

\subsection{Optimization of the effective nodal potential}
The trial wave-function is often constructed with non-interacting
orbitals derived from an effective potential [see Eq.
(\ref{eq:Heff})], e.g. from Kohn-Sham DFT. For the moment we will
assume that $\tilde{\Psi}_T({\bf R})$ is given in the single
determinant Slater-Jastrow form: $\tilde{\Psi}_T({\bf R}) =
e^{\tilde{J}({\bf R})} \tilde{\Phi}_T({\bf R})$ (this derivation is
extended to multiple determinants or pfaffians in
\ref{ssc:opmultideterminants}). However, for now, we assume that the
node {\it can move} within all the non-interacting $v$-representable
set, which is a less restrictive condition than the fixed-node
approximation but implies accepting an error if $S({\bf R})$ is not
$v$-representable.

In Ref.~\onlinecite{rosetta} we showed that, if the trial wave
function depends on non-interacting orbitals in an effective potential
[as in Eq.~(\ref{eq:Heff})], the effective potential $ \bar{V}\left(
  {\bf r} \right)$ required to retain a given property is a function
of the cost function $K$.  To simplify formulae, discussion and
notation we assume here that all wave-functions are real. The potential can
be obtained by adding recursively the following correction:

\begin{equation}
\label{eq:vK}
d V_{K}({\bf r} )  =  -
\epsilon \sum_{\nu}^{o}\int \!\!
 {\bf dr^{\prime}} \!\!
\frac{\delta K} {\delta \phi_{\nu}
\left( {\bf r^{\prime} } 
\right)
}
\frac{\delta \phi_{\nu} \left( {\bf r^{\prime}}\right) }
{\delta V_K \left( {\bf r}  \right) } .
\end{equation}
where $\epsilon$ is adjusted during the optimization. Replacing $K$ by
$K_{DMC}$ we get
\begin{equation}
\label{eq:dKDMC}
\frac{\delta K_{DMC}} {\delta \phi_{\nu} \left( {\bf r^{\prime} }\right ) }
 = \int {\bf dR} W({\bf R}) e^{\tilde{J}(\bf R)}
\frac{\delta \tilde{\Phi}_T({\bf R})}
{ \delta \phi_{\nu} \left( {\bf r^{\prime}  } \right ) }
\end{equation}
where
\begin{equation}
W({\bf R}) = 
\frac{ \delta K_{DMC} }{\delta \tilde{\Psi}_T({\bf R})} \; , \nonumber
\end{equation}
for which we obtain
\begin{eqnarray}
\label{eq:wr}
W({\bf R}) &= &
\frac{
2 \Delta({\bf R})   \alpha({\bf R})
 - \Delta({\bf R})^2}
{|\alpha({\bf R})|^2}  \mathrm{sign}(\alpha({\bf R}))
  \nonumber \\
&\times& \left [ 1- \alpha({\bf R} ) \right ]
\mu \; \Psi_T({\bf R}) \theta \left( f({\bf R}) - \eta \right )
,
\end{eqnarray}
with $\Delta({\bf R})= f({\bf R})-\alpha({\bf R})$.
Within first order perturbation theory
\begin{equation}
\label{eq:linearresp}
\frac{\delta \phi_{\nu} \left( {\bf r^{\prime} }\right) }
{\delta V_K \left( {\bf r}  \right)} = \sum_{n}^{u}
\frac{\phi_{n}({\bf r}) \phi_{\nu}({\bf r})}
     {\varepsilon_{\nu}-\varepsilon_n}
\phi_{n} \left( {\bf r^{\prime} }\right) \;.
\end{equation}
Replacing Eq.~(\ref{eq:dKDMC}) and Eq.~(\ref{eq:linearresp}) in
Eq.~(\ref{eq:vK}), we find
 \begin{eqnarray}
\label{eq:vDMC}
d V_{K_{DMC}}({\bf r} )  &=&
\epsilon
\sum_{\nu}^{o}
\sum_{n}^{u}
\frac{\phi_{n}({\bf r}) \phi_{\nu}({\bf r})}
     {\varepsilon_{\nu}-\varepsilon_n}  \beta^n_{\nu} \; , \\
\label{eq:betn}
  \beta^n_{\nu}   &=& \int {\bf dR} W({\bf R}) e^{\tilde{J}(\bf R)}
\tilde{\Phi}_{T,\nu}^{n}({\bf R}) \;.
\end{eqnarray}

In Eqs.~(\ref{eq:vK}), (\ref{eq:linearresp}), and (\ref{eq:vDMC}) we used
$\sum_{\nu}^{o}$ ( $\sum_{n}^{u}$ ) to define sums over occupied (unoccupied)
states. In turn in Eq.~(\ref{eq:betn}) $\tilde{\Phi}_{T,\nu}^{n}({\bf R})$
means replacing the occupied state $\phi_{\nu}$ by $\phi_{n}$ which results
from combining  the cofactors of $\phi_{\nu} \left( {\bf r^{\prime} } \right )$
[ $ \frac{\delta \tilde{\Phi}({\bf R})} { \delta \phi_{\nu}
\left( {\bf r^{\prime} } \right )} $ ] in Eq.~(\ref{eq:dKDMC})
with $\phi_{n} \left( {\bf r^{\prime}}\right)$ in Eq.~(\ref{eq:linearresp}).
The first factor in function $W({\bf R})$ [Eq.~(\ref{eq:wr})] is
obtained from the derivative of the cost function Eq.~(\ref{eq:KDMC}) with
respect to $\alpha({\bf R})$ [ignoring contributions coming from the
discontinuities of $|x|$ since the Heaviside function in Eq.
(\ref{eq:KDMC}) is zero near the nodes]. The second factor in $W({\bf R})$
results from the derivative of $\alpha({\bf R})$ with respect
to $\tilde{\Psi}_T({\bf R})$. [note that $\mu $ is also dependent on
$\tilde{\Psi}_T({\bf R})$ ]

\subsection{Optimization of the Jastrow factor within DMC}

We argued in the previous section that an optimal Jastrow factor can be
used to reduce the number of determinants in the multi-determinant expansion.
Optimizing the Jastrow factor is important to limit the exponential cost of
the CI expansion because, while the Jastrow factor cannot influence the nodes,
it can reduce the burden of correcting the probability density from any value
given by a Slater determinant (see Eq.~(\ref{eq:getJastrow})). Accordingly, if
the Jastrow factor is optimized, the antisymmetric part of the wave-function is
free to search for the nodes. Often the $\tilde{J}({\bf R})$ is
dependent on a set of parameters $\gamma_n$. The value of the cost
function (Eq.~\ref{eq:KDMC}) is also affected by the Jastrow factor
$e^{\tilde{J}(\bf R)}$. Thus the gradient of the cost function with
respect to an arbitrary change in $e^{\tilde{J}(\bf R)}$ can be obtained
within DMC via
\begin{equation}
\label{eq:opJ}
\frac{d K_{DMC}}{d \gamma_n} = \int {\bf dR} W({\bf R}) e^{\tilde{J}(\bf R)}
\tilde{\Phi}_T({\bf R} ) \frac{d \tilde{J}({\bf R)}} {d \gamma_n} .
\end{equation}

\subsection{Discussion}
Note at this point that (1) both the coefficients $\beta_{\nu}^n$ and
$\gamma_n$ are integrals of the function $W({\bf R})$ which is only
dependent on the particular form of the cost function selected in
Eq.~(\ref{eq:KDMC}) and a representation of the walkers distribution
$f({\bf R})$.

(2) The function $f({\bf R})$ is an essential component of 
$W({\bf  R})$ that can be obtained from the DMC run using
Eqs.~(\ref{eq:expaneta}) and (\ref{eq:sampcoeff}) or sampled directly
by binning.\cite{binnote}

(3) Provided that $f({\bf R})$ is known, a distribution of
configurations ${\bf R}_i$ with probability $|W({\bf R})|$ can be
generated with the Metropolis algorithm.  All integrals of the form
$\int {{\bf dR}}g({\bf R}) W({\bf R})$ involved in
Eqs.~(\ref{eq:vDMC}) and (\ref{eq:opJ}) can be evaluated in a single
correlated sampling step as $\sum_i \mathrm{sign}[W({\bf R}_i)] g({\bf
  R}_i)$ using points ${\bf R}_i$ drawn from the probability
distribution defined by the absolute value of $W({\bf R})$.

(4) In most methods, the Jastrow parameters $\gamma_n$ are optimized
within a variational Monte Carlo approach (either minimizing the total
energy or the energy variance). Here we optimize them within a DMC
run. The role of the Jastrow factor within this approach, is
different.  Its role instead is to correct the trial wave-function
$\tilde{\Phi}_T({\bf R})$ to match $\Phi_{FN}({\bf R})$.  The
optimization of the Jastrow parameters with Eq.~(\ref{eq:opJ}) only
ensures that the cost-function Eq.~(\ref{eq:KDMC}) is minimum.
Optimization of the Jastrow factor is required to allow the
antisymmetric part of the wave-function to move the nodes while the
Jastrow factor takes care of the symmetric contribution. However, if
the variational freedom of the Jastrow factor or the statistics are
limited, the minimization of Eq.~(\ref{eq:KDMC}) does not necessarily
imply a minimum in the VMC energy or its variance: the variance of the
local energy might rise. In those cases the Jastrow factor must be
optimized twice: first when the potential is optimized and second
during a VMC variance minimization before a collection DMC run.

Finally, (5)
note that $\tilde{\Psi}_T({\bf R})$ and $\Psi_T({\bf R})$ have
different Jastrow factors ($\Psi_T({\bf R})$ is kept fixed during the
cost function optimization steps).

\subsection{Optimization of multi-determinant wave-functions}
\label{ssc:opmultideterminants}
The multi-determinant expansion obtained in this
subsection is different from the one obtained in Section
\ref{sc:multideterminants}. In Section \ref{sc:multideterminants} we
found a multi-determinant expression of $\Psi_{FN}({\bf R})$ in a given
non-interacting orbital basis set for a given fixed Jastrow factor. Here
we optimize the Jastrow factor and the non-interacting basis to match
$\Psi_{FN}({\bf R})$ within a prescribed small number of determinants.

If we restrict the search to pure-state non-interacting
$v$-representable nodes, the minimum energy $E_{DMC}$ will be larger
than the true ground-state energy $E[\rho({\bf r})]$, because of the
upper-bound theorem, unless $S({\bf R})$ is $v$-representable.
In DMC the $v$-representability constraint is not required and can be
partially removed by including multi-determinants in $\Phi_T({\bf R})$
giving more variational freedom to the nodes.

Note that if we express $\tilde{\Psi}_T({\bf R})$ as a multi-determinant
expansion of the form
\begin{equation}
\label{apeq:multi}
\tilde{\Psi}_T({\bf R})= e^{J({\bf R})} \sum_k \alpha_k \tilde{\Phi}_k({\bf R}),
\end{equation}
an equivalent expression for wave-function optimization can be found.
The sum over occupied (unoccupied) levels in Eq.~(\ref{eq:vK}) must be
extended to every orbital that is occupied (unoccupied) in
$\tilde{\Phi}_k({\bf R})$. Also, it is easy to prove that the only
change in Eq.~(\ref{eq:vDMC}) required is in the values of the
$\beta^n_{\nu}$ which must be replaced by

\begin{eqnarray}
\beta^n_{\nu}   &=& \int {\bf dR} W({\bf R}) e^{\tilde{J}(\bf R)}
\sum_k \alpha_k c_n^{\dag} c_{\nu} \tilde{\Phi}_{k}({\bf R} )  \;,
\end{eqnarray}
where the operators $c_n^{\dag}$ and $c_{\nu}$ change, when possible,
the single particle state $\phi_{\nu}$ with $\phi_{n}$ in the Slater
determinant $\tilde{\Phi}_{k}({\bf R})$; and give zero if $\phi_{\nu}$
is not included or $\phi_{n}$ is already occupied. The function
$W({\bf R})$ is still given by Eq.~(\ref{eq:wr}). The coefficients
$\alpha_k$ can be optimized using the
following expression
\begin{equation}
\label{apeq:opak}
\frac{d K_{DMC}}{d \alpha_k} =
\int {\bf dR} W({\bf R}) e^{\tilde{J}(\bf R)}
\tilde{\Phi}_k({\bf R} )  \; .
\end{equation}

\section{Model system tests}
\label{sc:model}
In this section, to demonstrate the methods described above, we solve
a simple yet non-trivial interacting model as a function of the
interacting potential strength and shape. We then test a simple
version of the algorithm described in Section
\ref{sc:multideterminants}. Subsequently, we replace the model
interaction by a realistic Coulomb interaction. Finally, in subsection
\ref{ssc:model2} we optimize the wave-functions by obtaining the effective
nodal potential, as described in Section \ref{sc:singledeterminant}.

\subsection{A model interacting ground state}
\label{ssc:CI}
For illustrative purposes we choose the same problem studied in
Ref.~\onlinecite{rosetta} where we derived the existence of an
effective potential for the wave-function nodes.  Briefly, we solve
the ground state of two spin-less electrons moving in a two
dimensional square of side length $1$ with a repulsive interaction
potential of the form\cite{units} $V({\bf r},{\bf r^{\prime}}) = 8
\pi^2 \gamma \cos{[\alpha \pi(x-x^{\prime})]}\cos{[\alpha
  \pi(y-y^{\prime})]}$.  In this paper we show results for $\alpha= 1/
\pi$ and $\gamma = 4$.  With this choice of parameters the system is
in the highly correlated regime, because the matrix element of the
interaction potential between the non-interacting ground state and
first excited state is larger than the non-interacting energy
difference. We expand the many-body wave-function in a full CI
expansion of Slater determinants with the same symmetry as the ground
state. The ground state is degenerate because there are only two
electrons. We choose one of the ground-state wave-functions according
to the $D_2$ subgroup of the $D_4$ symmetry of the Hamiltonian. For
more details see Ref. \onlinecite{rosetta}.
\begin{figure}
\includegraphics[width=1.00\linewidth,clip=true]{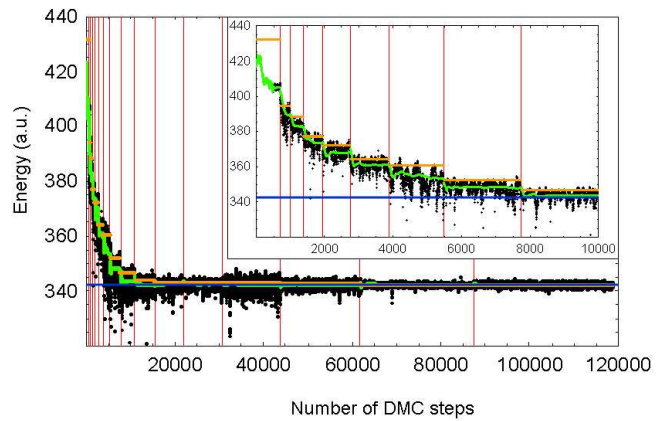}
\caption{(Color online) Self-healed DMC run obtained using the method
  described in Section \ref{sc:multideterminants}. Black points denote
  the average value of the local energy for each DMC step. Green
  points mark the reference energy used for population control. Orange
  lines mark the average energy of the trial wave-function. The
  horizontal blue line marks the energy of the ground state in the
  full CI calculation. Vertical lines mark the steps when the
  coefficients of wave-function are updated. Inset: Detail of the DMC
  run for the first 10000 steps (same conventions as in the main
  figure).}
\label{fg:dmcCIrun}
\end{figure}
From the full CI calculation we obtain a nearly exact expression of the
ground state $\Psi({\bf R})= \sum_n a_n \Phi_n({\bf R})$.

\subsection{Projection of the DMC fixed-node wave-function on a
multi-determinant expansion}
\label{ssc:model1}

In order to facilitate the comparison with the full CI results, we
sample the mixed-estimator density with the projectors $\xi_n({\bf R})$
constructed using the same basis functions of the CI expansion.
For the same reason, we utilized no Jastrow function ($J=0$ in Eq.~\ref{eq:xi}).

An initial trial wave-function must be selected. While the
non-interacting solution has very good nodes,\cite{rosetta} we
intentionally chose a poor initial trial wave-function in order to test the
strength of the multi-determinant method described in Section
\ref{sc:multideterminants}. The worst case scenario is when the trial
wave-function is orthogonal to the exact ground state. If the exact
ground state is not included in the trial wave-function, a projector
method such as the standard DMC algorithm {\it cannot} yield the exact
ground-state energy. Accordingly, to test the method, we chose for
this example $\lambda_1 = a_3$, $\lambda_3=- a_1$ and $\lambda_n = 0$
for all remaining $n$.\cite{basisnote} Expanding, $\Psi_{FN}({\bf R})$ with
these $\lambda_n$ and replacing it in Eq.~(\ref{eq:xi}) we obtain the
projectors $\xi_n({\bf R})$. Next we obtained new values $\lambda_n$
sampling Eq.~(\ref{eq:sampcoeff}) every autocorrelation time. After
many configurations are sampled, we construct a new trial
wave-function with the new $\lambda_n$. We only include in the
wave-function the coefficients that satisfied the condition
$|\bar{\lambda}_n| > 4 \frac{\langle\tilde{\sigma}_n\rangle}{\sqrt{N_c-1}}$,
i.e. that the coefficients are well determined according to this empirical
threshold. Note that because the multi-determinants are solutions of a
non-interacting problem, they will tend to have more nodes as their
energy increases. 
Accordingly, high energy components of the wave-function will have
smaller coefficients (${\lambda}_n$) in absolute value as compared
with the error ($\tilde{\sigma}_n$).
 As a consequence, this acceptance threshold
removes the contribution of the high energy components which implies
that the resulting wave-function will be smoother than
$\Psi_{FN}({\bf R})$ without the kinks at the nodes. This process is the
core of a more complex algorithm we propose for larger systems that is
explained in Section \ref{sc:algorithm} (see steps 3 and 4).

\begin{figure}
\includegraphics[width=1.\linewidth,clip=true]{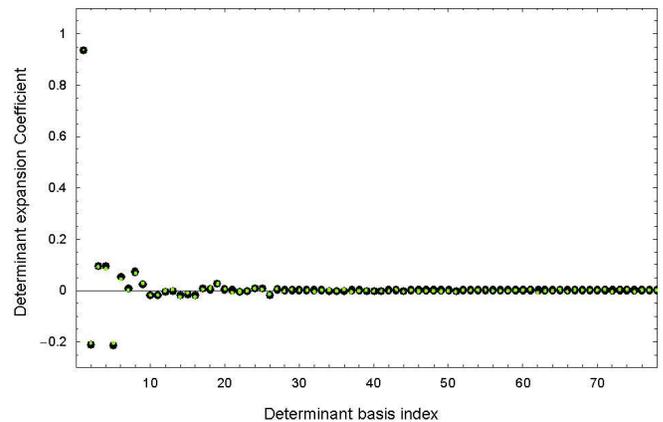}
\caption{(Color online) Values of the coefficients of the multi-determinant
  expansion (small green circles) as compared with a full CI
  calculation (large black circles). The DMC statistical errors of the
  coefficients is equal to the radius of the green circles. }
\label{fg:dmccoeff}
\end{figure}

\begin{figure}
\includegraphics[width=1.\linewidth,clip=true]{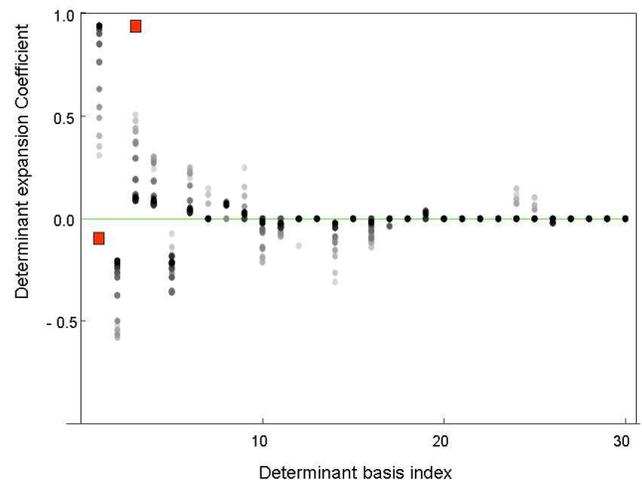}
\caption{(Color online) Change in the values of the multi-determinant
  expansion as the DMC self-healing algorithm progresses. Light gray
  colors denote older coefficients while darker ones denote more
  converged results. The initial non-zero coefficients are highlighted
  in red squares.}
\label{fg:dmccoeffb}
\end{figure}

\begin{figure}
\includegraphics[width=0.85\linewidth,clip=true]{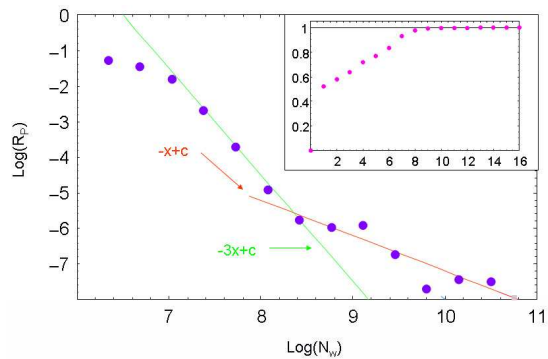}
\caption{(Color online) Logarithm of the residual projection $R_P$ [see Eq.~(\ref{eq:rp})] as
  a function of the total weighted number of configurations along the
  complete run $N_w$. The lines are a guide to the eye. Inset:
  projection of the DMC self-healed wave-function onto the full CI
  ground state as function of the logarithm of $N_w$.}
\label{fg:dmcproj}
\end{figure}

The result of this iterative approach is summarized in
Figs.~\ref{fg:dmcCIrun}, \ref{fg:dmccoeff}, \ref{fg:dmccoeffb} and
\ref{fg:dmcproj}. In Fig.~\ref{fg:dmcCIrun} we show the average of the
local energy $E_L$ (black dots) and the best estimator for the energy
$E_{best}$ (green dots)\cite{umrigar93} as a function of the number of
DMC steps. The average energy of the trial wave-function
$\bar{E}=\langle \Psi_T|H|\Psi_T\rangle/\langle\Psi_T|\Psi_T\rangle$
(orange) is also given for comparison. The run was carried out for a
targeted population of $200$ walkers. The exact full CI result is
given by the blue line.  There is a dramatic decrease of $E_L$,
$E_{best}$ and $\bar{E}$ as the trial wave-function is updated, and
all these values converge to the full CI result. Similar results are
obtained with different starting points and interaction strengths. The
only limiting factor to reaching the exact CI results appears to be
the iteration time.  The reduction in the energy variance can be seen
in Fig.~\ref{fg:dmcCIrun} where the fluctuations in the local energy
decrease as the run continues.

In Fig.~\ref{fg:dmccoeff} we show a plot of the values of the full CI
coefficients as a function of the coefficient index compared with the
average values obtained from the optimized trial wave-function and a
final DMC run using Eq.~(\ref{eq:sampcoeff}). The coefficients are
ordered with increasing non-interacting energy. The error bars of the
coefficient are also given. The figure shows that a wave-function
expansion with the quality of a CI expansion can be obtained with DMC.
Note that (i) knowledge of the ground-state wave-function allows for
the calculation of any other observable with an error bar that can be
obtained from the error bars of the expansion coefficients.
(ii) The same wave-function
could be expressed with a smaller number of determinants if a Jastrow
factor had been used.

In Fig.~\ref{fg:dmccoeffb} we show the evolution of the values of the full
CI coefficients as a the algorithm progresses starting from a trial 
wave-function orthogonal to the ground state. 

The improved quality of the DMC optimized trial wave-function is
also evident in Fig.~\ref{fg:dmcproj}. We plot the
logarithm of the residual projection
\begin{equation}
\label{eq:rp}
R_P=\log{[1-\langle \Psi|\Psi_T\rangle/(|\Psi||\Psi_T|)]}
\end{equation}
on the ``exact'' CI ground state as a function of the logarithm of the total
weighted number of configurations along the complete run $N_w$.
Remarkably, the error of the wave-function projection has decreased to
$e^{-8}$ starting from 1. By noting that $|\Psi_T\rangle = |\Psi \rangle+
|\delta \Psi_\perp \rangle $, where $|\delta \Psi_\perp \rangle$ is the
difference between the ground-state $|\Psi \rangle$ and the trial wave
function $|\Psi_T \rangle$ we get
\begin{equation}
\label{eq:rpapprox}
R_P \simeq 2 \log \left[|\delta \Psi_\perp|/\sqrt{2} \right] \;.
\end{equation}
We can see that for a significant section of the run $R_P \sim
1/N_w^3$, where $N_w$ is the total number of weighted configurations
of the run. This means that the magnitude of the error in the trial
function decays with a {\it faster} exponent than $1/\sqrt{N_w}$
($3/2$). This is surprising because if we had provided the exact
ground state as trial wave-function, the error after finite sampling
would have scaled as $|\delta \Psi_\perp| \sim 1/\sqrt{N_w}$, 
which replaced in Eq.~(\ref{eq:rpapprox}) gives $R_P \simeq 1/\sqrt{N_w}$.  This
faster exponent, in a section of the plot, is a direct consequence of
the fact that both the quality of the trial wave-function and the
statistics have improved. This is another indication that the nodes
continue to improve along the run. For the final part of the graph
(the last three points), however, $R_P$  scales as $1/\sqrt{N_w}$. 
This possibly signals that after the nodal structure is improved to a critical 
distance from
the exact ground state, the statistical error in the determination of
the coefficients and not a small fluctuation in the nodal structure,
is the limiting factor for this algorithm. We believe that a final
$1/\sqrt{N_w}$ scaling of $R_P$ signals also that the overall nodal
structure of the solution is correct and only small fluctuations of
the coefficients are responsible for the small fluctuations from the
exact node.

Since a direct sampling of the fixed-node wave-function
(Eq.~(\ref{eq:sampcoeff})) aims to reproduce the fixed-node solution, a
single DMC run cannot improve the nodes. Only by iterating with different
trial wave-functions can the nodes be improved. In particular, if an
infinite number of configurations were used, the nodes would not change.
In practice however, we find that for a finite sample, the error in the
wave-function coefficients plays a positive role. Errors act as random
fluctuations in a simulated annealing algorithm. These fluctuations are
reinforced\cite{fn:reinforced} or discarded in subsequent iterations. This allows the nodal
error to be systematically reduced to the point that trial wave-functions
with 0.9995 projections on the full CI ground state can be found starting
from a trial wave-function initially orthogonal to the ground state. Since
poor nodes are associated with discontinuities in the derivative of
$\Psi_{FN}({\bf R})$ at the nodal surface, and consequently an increase in
the kinetic energy, it is also convenient at first to initially limit the
number of configurations sampled (including first the ones that cost
less non-interacting energy).

We recognize that the current work does not address the suitability and
convergence of this method of relying on random fluctuations for systems
with large numbers of electrons; this will be the subject of later studies.

\subsection{Coulomb potential results and discussion}
The use of a simplified electron-electron interaction facilitates the
CI calculations and the validation of the optimization method
described in Section \ref{sc:multideterminants}. However,
it is also important to test the convergence and stability of the
method with a realistic Coulomb interaction. Note that in two
dimensions (2D) the correlations are enhanced as compared with three
dimensions (3D) while the nodal surface remains non-trivial.

We tested the stability of the algorithm by replacing the interaction
potential with:\cite{units} $V({\bf r},{\bf r^{\prime}}) = 20 \pi^2 /|{\bf
  r-r\prime}|$. Since the length of the square box side is 1,
the difference in kinetic energy between the non-interacting ground
state and the first-excited state is $3 \pi^2$. This choice of parameters
for the Coulomb potential placed the system in a strongly interacting
regime. To further increase the role of correlations and the difficulties
that the algorithm must overcome we did not included a Jastrow term, i.e. 
$J=0$. We also increased the chances of {\it failure} by setting the initial trial 
wave-function equal to the {\it first excited state} of the non-interacting 
system.
\begin{figure}
\includegraphics[width=0.85\linewidth,clip=true]{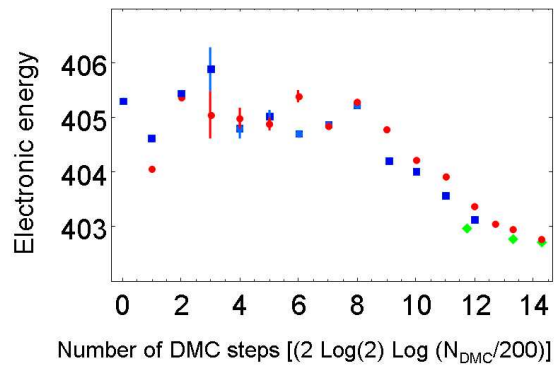}
\caption{(Color online) Energy of the DMC run as a function of the number
  of DMC steps used to gather statistical data of the wave-function in
  the previous block. The statistical error bars for the first three points
  on the left were not calculated. The statistical error bars of the points
  on the right were smaller than the size of the symbols. Blue squares 
  denote calculations starting from a bad trial wave-function, while
  the red circles mark the results obtained from an initial trial
  wave-function corresponding to the best blue square on the right (see
  text). Green diamonds were generated starting from the best red circle.}
\label{fg:Coulombenvar}
\end{figure}

In Fig.~\ref{fg:Coulombenvar} we show the evolution of the average
of the local energy for each DMC optimization block as a function of
the number of DMC steps in each optimization block $N_{DMC}$. Data for
Eq.~(\ref{eq:sampcoeff}) is accumulated every $100$ DMC steps. As in
the case of the model Hamiltonian, we increase $N_{DMC}$ in each
optimization as $N_{DMC}=200\times 2^{n_b/2}$ where $n_b$ is the
total number of blocks. With this choice we can expect the error bar in the
energy and in the coefficient $\lambda_n$ of the multi-determinant
expansion Eq.~(\ref{eq:expan}) to be reduced a factor $1/2$ after four 
successive blocks. Note that during each DMC run not only the local energy is
sampled but also the values of the projectors $\xi({\bf R})$ used to
construct the expansion of the trial wave-function of the next
point on the right with Eq.~(\ref{eq:sampcoeff}).

The blue squares in Fig.~\ref{fg:Coulombenvar} show the progression
in average DMC energy starting from the first excited state. The initial
energy is above 420 compared with the fully converged energy of
402.718 $\pm$ 0.008. Even starting from such a bad initial trial wave
function, our method is able to improve in the second block after only
accumulating $\approx 400$ configurations. In contrast, the red circles in
Fig.~\ref{fg:Coulombenvar} denote the results obtained with an initial
trial wave-function constructed with data collected with the right most
blue square, a very good initial trial wave-function.

As the optimization process is repeated, the average DMC energy
fluctuates. Since the coefficients carry a statistical
error, the wave-function is not the same from one block to the other
and neither is the nodal error. There is a shift from one iteration to
the next which is sometimes larger than the error bar in the energy.
The energy and the variance can fluctuate and locally increase.
However, as the statistics improve, fluctuations in the coefficients
decrease. The statistical errors play the role of a thermal noise in
the coefficient expansion. Improved statistics correspond to reduced
temperatures in simulated annealing. Note that, initially, the average
DMC energy from the very poor trial wave-function decreases (blue
squares) as the algorithm progresses, while the energy of the average
DMC energy from the good trial wave-function (red circles) actually
increases.  This is because when the statistics are poor the errors in
the coefficient expansion allows improvement of a bad trial wave
function but spoil a good quality one. Figure \ref{fg:Coulombenvar}
shows that, as the algorithm progresses and improved statistics are
obtained, the quality of the solution becomes independent the initial
trial wave-function. Note that for intermediate blocks the DMC energy becomes
flat, signaling that the the statistics are not enough to reduce the nodal
error, but are sufficient to stop deterioration of  the wave-function.

Repeating the algorithm iteratively leads to an incremental improvement
in the statistics which results in a clear reduction of the DMC energy
beyond the error bar of the preceding calculations. The DMC energy and
the energy variance are reduced systematically which is a clear
indication of the reduction of nodal errors and improvement in the
overall quality of the wave-function.  The ground-state energy
obtained after 240000 accumulated DMC iterations is 402.718 $\pm$
0.008.

\begin{figure}
\includegraphics[width=0.85\linewidth,clip=true]{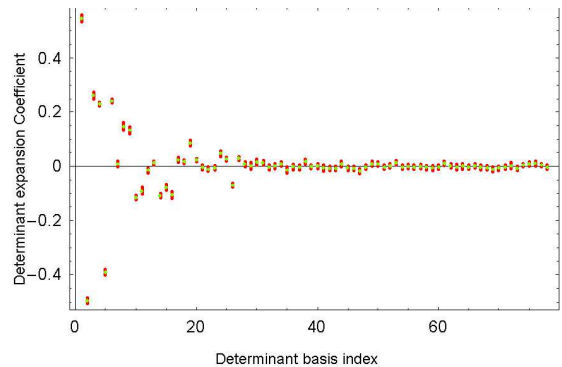}
\caption{(Color online) Values of the coefficients of the multi-determinant expansion
  (small green circles) obtained from the DMC run for two electrons in
  a square box with a Coulomb interaction in the highly correlated
  limit. The statistical errors in the values of the coefficients are
  equal to the size of the red bar. }
\label{fg:Coulombcoeff}
\end{figure}
In Fig.~\ref{fg:Coulombcoeff} we show the values of the coefficients
of the multi-determinant expansion as obtained with Eq.
(\ref{eq:sampcoeff}) corresponding to the right-most blue point in Fig.
\ref{fg:Coulombenvar}. Note that since no Jastrow factor is used and the
interaction potential includes a singularity at ${\bf r}={\bf r^\prime}$,
the number of coefficients with significant value is much
larger the model interaction described earlier. The final reduction of
nodal errors shown in the final steps of Fig.~\ref{fg:Coulombenvar} is
associated with subtle variations of the coefficients.

If the Jastrow factor is set to one, the density takes a simple form
(Eq.~(\ref{eq:density})) in terms of the single-particle orbitals
$\phi_n({\bf r})$. Knowledge of this density allows the calculation of
the Kohn-Sham potential as explained in Ref.~\onlinecite{rosetta} (see below)
and suggests an alternative route for calculation of forces by applying\cite{payne92} 
the Hellmann-Feynman theorem directly to the Kohn-Sham total energy instead 
of the usual statistical sampling.\cite{badinski08,filippi00prbr} 
The DMC density can be obtained in terms of the single-particle orbitals 
with the following equation:\cite{QTS}
\begin{equation}
\label{eq:density}
\rho({\bf r}) = \sum_{n,\nu} \phi_n^*({\bf r})\phi_{\nu}({\bf r})
\sum_{k,l} \lambda_k^* \lambda_l \langle \Phi_k|c_n^{\dag} c_{\nu}|\Phi_l
\rangle \; .
\end{equation}
Note in Eq.~(\ref{eq:density}) that all the matrix elements
$\langle \Phi_k|c_n^{\dag} c_{\nu}|\Phi_l \rangle$ corresponding to states
that differ in more than one electron hole pair, do not contribute to the
ground-state density.
\begin{figure}
\includegraphics[width=0.85\linewidth,clip=true]{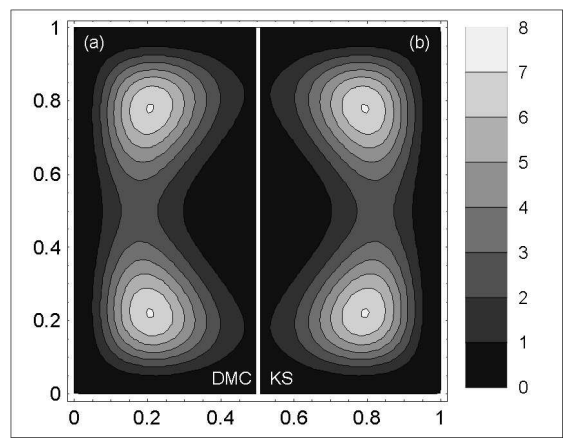}
\caption{Density of the ground state of two spin-less electrons with
  Coulomb interaction in a square box. We choose one of the two
  degenerate ground states, reducing the symmetry of the density
  to $D_2$. a) Left side of the density of the many-body
  ground state constructed with the converged coefficients shown in
  Fig.~\ref{fg:Coulombcoeff}. b) Kohn-Sham non-interacting density
  constructed as explained in Ref.~\onlinecite{rosetta}.}
\label{fg:Coulombden}
\end{figure}
\begin{figure}
\includegraphics[width=0.85\linewidth,clip=true]{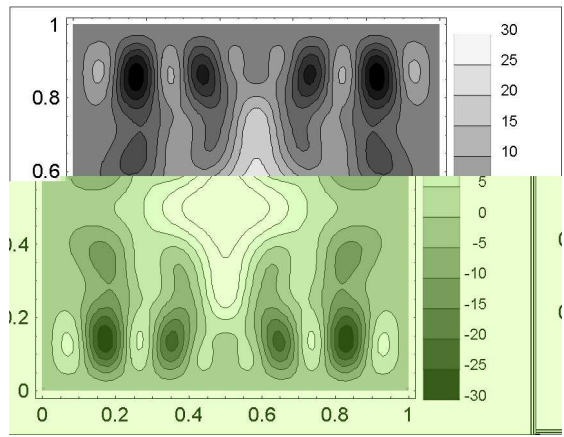}
\caption{Kohn-Sham potential for two spin-less electrons in a square box
  corresponding to the ground state of Figs.~\ref{fg:Coulombcoeff} and
  \ref{fg:Coulombden}. The potential was constructed using the methods
  explained in Ref.~\onlinecite{rosetta}.}
\label{fg:Coulombpot}
\end{figure}
In Fig.~\ref{fg:Coulombden}a we show the density corresponding to
the coefficients of Fig.~\ref{fg:Coulombcoeff} and in
Fig.~\ref{fg:Coulombden}b the non-interacting Kohn-Sham density
constructed using the methods explained in Ref.~\onlinecite{rosetta}.

In Fig.~\ref{fg:Coulombpot} we show the Kohn-Sham potential obtained
using the methods described in Ref.~\onlinecite{rosetta}. We minimized the
cost function in Eq.~(2) of Ref.~\onlinecite{rosetta} using 14 Fourier
components in the potential expansion. We believe that the sampled
oscillations in the Kohn-Sham potential carry some physical meaning.
Indeed, these oscillations are required in order to match the
non-interacting density in Fig.~\ref{fg:Coulombden}b to the
interacting self-healed DMC density in Fig.~\ref{fg:Coulombden}a.
However, since the density $\rho({\bf r})$ has an error
$\sigma_{\rho}({\bf r})$, there is also an error in the Kohn-Sham
potential. In linear response,\cite{rosetta} the error bar in the
potential $\sigma_{KS}({\bf r})$ (not shown) can be obtained in terms
of $\sigma_{\rho}({\bf r^{\prime}})$ and the inverse susceptibility
as
\begin{equation}
\label{eq:errorpot}
\sigma_{KS}({\bf r})  = \int  {\bf dr^{\prime}}
\sigma_{\rho}({\bf r^{\prime}})
\frac{\delta V \left( {\bf r^{\prime} } \right) }{\delta \rho
\left( {\bf r} \right) } \;\;\; .
\end{equation}

Since, we have removed degeneracies in the ground state by restricting
the symmetry of the wave-function, two potentials that give the same
density can only differ by a constant. We have obtained
from DMC not only the approximated DMC energy but also the derivative
of the total energy with respect to local fluctuations of the density.
Figures \ref{fg:Coulombden} and \ref{fg:Coulombpot} show that
this method can provide accurate benchmarks for the validation of DFT
approximations in the highly correlated regime.

\subsection{Model system effective nodal potential and Jastrow factor}
\label{ssc:model2}
To demonstrate that the effective nodal potential and Jastrow factor
can be obtained through sampling in DMC, in this section we determine
these quantities for a model corresponding to two electrons in a
square box with Coulomb interactions. An additional goal is to show
that a complex (multi-determinant) wave-function can potentially be
replaced by a simpler one while retaining the same nodal structure.

The results below correspond to a trial wave-function represented using
the multi-determinant expansion shown in Fig.~\ref{fg:Coulombcoeff}.
While for larger dimensional systems the integrals can be performed
more efficiently using a stochastic approach, in this case the
probability densities were binned numerically over a grid of fifteen
bins in all four dimensions.  Approximately, $7.2 \times 10^5$
weighted\cite{weighnote} configurations were collected.

The one-body and two-body Jastrow factors were simply written as a
Fourier expansion and their coefficients were minimized with an
accelerated steepest decent algorithm using Eq.~(\ref{eq:opJ}). The
antisymmetric part of the wave-function was given by a single
determinant corresponding to the ground-state solution of a
non-interacting effective potential. The effective interactive
potential was expressed as a sum of cosine functions and optimized as
explained in Ref.~\onlinecite{rosetta}. The Jastrow factors and the 
potentials can be optimized at the same time. However, since we
wanted the Jastrow factor to carry most of the load in the optimization of
the symmetric corrections to the probability density, the potential
was optimized only every third iteration that the Jastrow factor
was optimized.

\begin{figure}
\includegraphics[width=0.85\linewidth,clip=true]{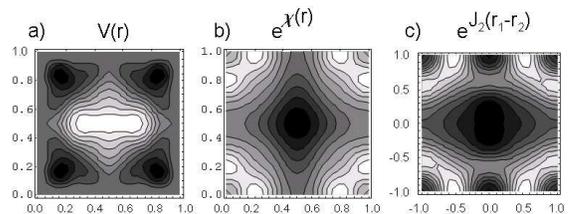}
\caption{a) Effective nodal potential, b) one-body Jastrow, and c)
  two-body Jastrow factors obtained by minimizing Eq.~(\ref{eq:KDMC}),
  in which the multi-determinant expansion of Fig.~\ref{fg:Coulombcoeff}
  has been replaced by a single determinant function.}
\label{fg:optimization}
\end{figure}

The resulting potential, and Jastrow factors are shown in
Fig.~\ref{fg:optimization}. The value of the cost-function was
reduced an order of magnitude from starting with the non-interacting
ground state with zero effective potential. The effective potential
resulting from this minimization procedure is an example of the nodal
potential predicted in Ref.~\onlinecite{rosetta}.

We also performed tests of this optimization algorithm using the
model interaction discussed in Subsection \ref{ssc:model1}.
In this case the nodal structure of the wave-function was also improved
(as signaled by a reduction of the average DMC energy below the error
bar of the preceding calculation).

\section{Summary of improved self-healing DMC algorithm}
\label{sc:algorithm}
It is clear from previous sections that an effective wave-function optimization
algorithm can be constructed solely on the basis of iteratively updating $\Psi_T$
by the multi-determinant expansion of $\Psi_{FN}$. An example of this algorithm
applied to a soluble model is presented in Subsection
\ref{ssc:model1}. However, multi-determinant expansions in DMC are
computationally very expensive in large or continuum system, since the required
number of determinants to reach a given accuracy will in general grow
combinatorially. The method developed in Section \ref{sc:singledeterminant}
to optimize a single Slater determinant becomes very attractive. (Results of
the application of this method were shown in Subsection \ref{ssc:model2}). For
large systems, the number of multi-determinants must be kept to a minimum and
the two methods combined. Experimentation in small systems allows us to suggest
an algorithm that will be efficient in larger systems:
\begin{enumerate}
\item An initial trial trial wave-function  $\Phi_T({\bf R})$ is generated
using any fast method, e.g. an empirical screened pseudopotential\cite{wang95}
or a Thomas-Fermi theory.

\item The Jastrow factor $J({\bf R})$ is optimized within VMC.

\item A DMC run is performed. The number of configurations $N_c$ sampled
is increased as this step is repeated. Statistically uncorrelated values of
$\xi_n({\bf R})$ and $\xi_n({\bf R})^2$  are accumulated.

\item The multi-determinant expansion of $f({\bf R})$ is constructed.
Only the terms that are significantly non-zero are included in the
expansion.

\item A distribution of configurations ${\bf R}_i$ with probability
$|W({\bf R})|$ is generated.  The gradients of $K_{DMC}$ with
respect to the effective nodal potential and the gradients of the
Jastrow factor coefficients are evaluated with Eqs.~(\ref{eq:vDMC}),
and (\ref{apeq:opak}). (Eventually the multi-determinant expansion
coefficients $\alpha_k$ can be included, see Subsection
\ref{ssc:opmultideterminants}.)

\item The effective potentials $V \left({\bf r} \right)$ and
$\tilde{J}({\bf R})$ are updated (eventually also the $\alpha_k$).
New single particle orbitals are constructed using Eq.~(\ref{eq:Heff}).
Therefore the single particle orbitals used to construct the Slater
determinants in the trial wave-function are now determined solely within DMC.

\item A new $\tilde{\Psi}_T({\bf R})$ is constructed. Steps (5-7) are
repeated until $\tilde{\Psi}_T({\bf R})$ does not change.

\item At this step we can choose to improve the scaling in large systems.
The single-particle
orbitals $\phi_n({\bf r})$ shared by all determinants in the expansion 
$\tilde{\Phi}_T({\bf R})$ 
can be transformed to non-orthogonal localized orbitals.\cite{reboredo05,alfe04}

\item The trial wave-function $\Phi_T({\bf R})$ is updated to
$\tilde{\Phi}_T({\bf R})$. Steps (2-9) are repeated until $\tilde{\Psi}_T({\bf R})$
and $E_{DMC}$ do not change.\cite{fn:change}

\end{enumerate}

Note that (i) the methods in Sections \ref{sc:multideterminants} and
\ref{sc:singledeterminant} are complementary. In Section
\ref{sc:multideterminants}, we find a representation of the fixed-node
ground state in a given basis. In Section \ref{sc:singledeterminant},
instead, we optimize and change the basis of the wave-functions so as
to reproduce the fixed-node ground-state wave-function with a minimum
number of Slater determinants. (ii) Only single configurations are
included in Eq.~(\ref{eq:vDMC}) but multiple configurations are
included in Eq.~(\ref{eq:expan}). (iii) We include a Jastrow function
in Eq.~(\ref{eq:expan}) to minimize the number of Slater determinants
required in the expansion. However, a final run with no Jastrow factor
included with the configuration interaction expansion might be useful
in order to obtain a pure expression of the ground-state density in
terms of the single particle orbitals. Atomic forces could be obtained
from this density.  Finally (iv) the method is, in principle, self
reliant: no DFT or HF are required.

\section{Summary}
\label{sc:discussion}
We have presented an algorithm for sampling the fixed-node many-body
wave-function in a single or multi-determinant expansion from a
diffusion quantum Monte Carlo (DMC) calculation within the importance
sampling technique. By combining this algorithm with a previously
developed method for constructing effective potentials targeted at
reproducing specific properties of the many-body
wave-function,\cite{rosetta} we presented an iterative algorithm that
improves the nodes of the trial/fixed-node wave-functions used in DMC.
Tests on a simple two electron model system confirm that this method
is able to improve the nodes and that, at least in the case of the
tested system, we find wave-functions and energies that exactly match fully
converged configuration interaction calculations.

We have proven that the nodes of the fixed-node wave-function improve
as compared with the trial wave-function if the kinks at the nodes are
locally smoothed out.  The algorithms presented take advantage of this
proof.  We have argued that if the kink at the node increases with the
``distance'' from the exact ground-state node to the trial
wave-function node, the algorithm should be stable against random
statistical fluctuations.  Proving this property in general might be
difficult and is beyond the scope of this article. Clearly, in the
absence of a proof, experimentation in larger systems is required.

While in the past, methods were used to obtain the fixed-node wave-function
(e.g Ref.~\onlinecite{bianchi96}), to our knowledge this is the
first time the fixed-node wave-function has been obtained through
importance sampling.  The availability of the fixed-node wave-function
provides routes to determine the exact Kohn-Sham potential, allowing
benchmark tests of density functionals in highly non-trivial and
inhomogeneous systems. It also seems likely that many of the wave
function optimization approaches
(e.g. Refs.~\onlinecite{filippi00,umrigar07,rios06,luchow07}) currently
applied within variational Monte Carlo can be recast in the
present scheme, making direct use of the fixed-node wave-function, and
likely obtaining improved results.

In ongoing work, we are continuing to develop these methods.
Applications to larger and more complex electronic systems will be
reported elsewhere.

\section*{Acknowledgments}
Research performed at the Materials Science and Technology Division
and the Center of Nanophase Material Sciences at Oak Ridge National
Laboratory sponsored the Division of Materials Sciences and the
Division of Scientific User Facilities U.S. Department of Energy.
This work performed under the auspices of the U.S. Department of Energy
by Lawrence Livermore National Laboratory under Contract DE-AC52-07NA27344.
The authors would like thank J. Kim for discussions and C. Umrigar for
clarifications related to the use of Eq.~(\ref{eq:gamma}).


\begin{thebibliography}{99}
\bibitem{anderson79} J. B. Anderson, Int. J. of Quantum Chem. {\bf 15}, 109
(1979).

\bibitem{ceperley80} D. M. Ceperley and B. J. Alder,
Phys. Rev. Lett. {\bf 45}, 566 (1980).

\bibitem{fn:energy} The energy of any wave-function $\Phi$ being 
$\langle \Phi|\hat\mathcal{H} |\Phi \rangle / \langle \Phi|\Phi \rangle$ 
for a given Hamiltonian $\hat\mathcal{H}$.

\bibitem{reynolds82} P. J. Reynolds, D. M. Ceperley, B. J. Alder, and W.
A. Lester, J. Chem. Phys. {\bf 77}, 5593 (1982).

\bibitem{bajdich05} M. Bajdich, L. Mitas, G. Drobny, and L. K. Wagner,
Phys. Rev. B {\bf 72}, 075131 (2005); L. Mitas, Phys. Rev. Lett. {\bf 96},
240402 (2006).

\bibitem{filippi00} C. Filippi and S. Fahy, J. Chem. Phys. {\bf 112}, 3523
(2000).

\bibitem{umrigar07} C. J. Umrigar, J. Toulouse, C. Filippi, S. Sorella, and
R. G. Hennig, Phys. Rev. Lett. {\bf 98}, 110201 (2007).

\bibitem{rios06} P. L\`{o}pez-Rios, A. Ma, N. D. Drummond, M. D. Towler, and
R. J. Needs, Phys. Rev. E {\bf 74}, 066701 (2006).

\bibitem{luchow07} A. L\"{u}chow, et. al., J. Chem. Phys. {\bf 126}, 144110
(2007).

\bibitem{kalos00} M. H. Kalos and F. Pederiva, Phys Rev. Lett. {\bf 85}, 3547
(2000).

\bibitem{ceperley84} D. M. Ceperley and B. J. Alder, J. Chem. Phys.
{\bf 81}, 5833 (1984).

\bibitem{zhang91} S. W. Zhang and M. H. Kalos, Phys. Rev. Lett. {\bf 67}, 3074
(1991).

\bibitem{beaudet} T. D. Beaudet, M. Casula, J. Kim, S. Sorella, and R. M.
Martin, J. Chem. Phys. {\bf 129}, 164711 (2008). 

\bibitem{toulouse08} J. Toulouse and C. J. Umrigar, J. Chem. Phys.
{\bf 128}, 174101 (2008).

\bibitem{williamson} A. J. Williamson, R. Q. Hood, and J. C. Grossman,
Phys. Rev. Lett. {\bf 87}, 246406 (2001).

\bibitem{reboredo05} F. A. Reboredo and A. J. Williamson, Phys. Rev. B
{\bf 71}, 121105(R) (2005).

\bibitem{alfe04} D. Alfe and M. J. Gillan, Phys. Rev. B  {\bf 70}, 161101(R)
(2004).

\bibitem{rosetta} F. A. Reboredo and P. R. C. Kent, Phys. Rev. B. {\bf 77},
245110 (2008).

\bibitem{tiago08jpc} M. L. Tiago, P. R. C. Kent, R. Q. Hood, and F. A. Reboredo,
J. Chem. Phys. {\bf 129}, 084311 (2008).

\bibitem{HLRbook} B. L. Hammond, W. A. Lester, Jr., and P. J. Reynolds
{\it Monte Carlo Methods in Ab Inition Quantum Chemistry} (World Scientific,
Singapore-New Jersey-London-Hong Kong, 1994).

\bibitem{shadow} S. Vitiello, K. Runge, and M. H. Kalos, Phys. Rev. Lett.
{\bf 60}, 1970 (1988).

\bibitem{fn:purediff} Equation~(\ref{eq:DMC}) can be thought of as a one-step 
($M=1$ in Eq.~(\ref{eq:propagator_m})) in a pure-diffusion DMC run using the small $\tau$ 
propagator of Eq.~(\ref{eq:delta}), in which the positive and negative walkers
have the initial distribution $\Psi_{FN}({\bf R^\prime})$ and are not constrained
by the nodes of $\Psi_{T}({\bf R})$. 
\bibitem{fn:bosons} Because of elemental group theory theorems, 
i) the product of two anti-symmetric functions is symmetric, ii) the product
of any symmetric operator with any anti-symmetric function results in
an anti-symmetric function (with no projection on any symmetric bosonic
function or any other representation of the symmetric group). Thus our
analytic derivation is true. Since we use an anti-symmetric representation
of the wave-function, our numerical method will {\it not} find bosonic states
since they are excluded from the search.\cite{bianchi93}
\bibitem{fn:reinforced} The nodes continue to
move statistically in the same direction in successive iterations.
\bibitem{fn:relnode} The ``release-node'' method \cite{ceperley80,HLRbook}
involves two fixed trial wave-functions and two populations of
walkers. Our method is very different since, instead, involves only one trial 
wave-function that is updated and a single population of walkers. In our method 
the walkers {\it never} cross the node: we move the node instead. In the so called 
release-node method the node is removed.
\bibitem{bianchi93}  R. Bianchi, D. Bressanini, P. Cremaschi, M. Mella, and
G. Morosi, J. Chem. Phys. {\bf 98}, 7204 (1993).

\bibitem{bianchi96} R. Bianchi, D. Bressanini, P. Cremaschi, M. Mella, and
G. Morosi, Int. J. of. Quant. Chem. {\bf 57}, 321 (1996).

\bibitem{us} F. A. Reboredo and C. R. Proetto, Phys. Rev. B {\bf 67}, 115325
(2003).

\bibitem{note} In Ref. \onlinecite{us} the projectors $\xi_n({\bf R})$ are
denoted as $\psi_i^{\sigma}(z)$.


\bibitem{umrigar93} C. J. Umrigar, M. P. Nightingale, and K. J. Runge,
J. Chem. Phys. {\bf 99}, 2865 (1993).

\bibitem{wood06} B. Wood and W. M. C. Foulkes, J. of Phys. C {\bf 18},
2305 (2006).

\bibitem{correa05} A. A. Correa, F. A. Reboredo, and C. A. Balseiro,
Phys. Rev. B {\bf 71}, 035418 (2005).

\bibitem{hohenberg} P. Hohenberg and W. Kohn, Phys. Rev. {\bf 136}, B864 (1964).

\bibitem{parr} R. G. Parr and W. Yang, {\it Density-Functional Theory
of Atoms and Molecules} (Oxford Science Publications, Oxford, 1989).

\bibitem{kohn} W. Kohn and L. J. Sham, Phys. Rev. {\bf 140}, A1133 (1965).

\bibitem{binnote} In section \ref{ssc:model2} we solve a case where it is more
efficient to bin $f({\bf R})$.

\bibitem{units} We define the energy unit to be $\hbar^2/(2 m)$.

\bibitem{basisnote} The basis functions in the CI expansion are ordered with increasing
non-interacting energy.

\bibitem{payne92} M.C. Payne, M.P. Teter, D.C. Allan, T.A. Arias,
and J.D. Joannopoulos Rev. Mod. Phys. {\bf 64}, 1045 (1992).

\bibitem{badinski08} A. Badinski and R. J. Needs, Phys. Rev. B {\bf 78},
035134 (2008).

\bibitem{filippi00prbr} C. Filippi and C. J. Umrigar, Phys. Rev. B {\bf 61},
R16291 (2000).

\bibitem{QTS} C. Kittel {\it Quantum Theory of Solids} (John Wiley \& Sons,
New York, 1987).

\bibitem{weighnote} The weight of the walkers goes from 2 to 1/2 and is
multiplied by the correction given in Eq.~(\ref{eq:gamma}).

\bibitem{wang95} L.-W. Wang and A. Zunger, Phys. Rev. B {\bf 51}, 17398 (1995).

\bibitem{fn:change} $\tilde \Psi$ must be converged twice:
first in order to fit the fixed-node solution which allows the node to
move, and second when the node no longer moves.
\end{thebibliography}
\end{document}